\documentclass[sigconf]{acmart}

\AtBeginDocument{%
  }

\copyrightyear{2023}
\acmYear{2023}
\setcopyright{rightsretained}
\acmConference[SSDBM 2023]{35th International Conference on Scientific and Statistical Database Management}{July 10--12, 2023}{Los Angeles, CA, USA}
\acmBooktitle{35th International Conference on Scientific and Statistical Database Management (SSDBM 2023), July 10--12, 2023, Los Angeles, CA, USA}
\acmDOI{10.1145/3603719.3603726}
\acmISBN{979-8-4007-0746-9/23/07}

\newcommand{\mybox}[1]
{
\vspace{2mm}
\noindent \hspace{-1mm} 
\setlength\fboxsep{1mm}
\fbox{\parbox{\dimexpr\linewidth-2\fboxsep-2\fboxrule}{\itshape #1}}
\vspace{0.2mm}
}

\usepackage{graphicx}
\usepackage[ruled, vlined, linesnumbered]{algorithm2e}
\DeclareUnicodeCharacter{0301}{\'{e}}

\newcommand{\para}[1]{\vspace{1mm}\noindent\textbf{#1.}}

\begin{document}

\title{Accelerating Machine Learning Queries with Linear Algebra Query Processing}

\author{Wenbo Sun}
\email{w.sun-2@tudelft.nl}
\affiliation{%
  \institution{Delft University of Technology}
  \city{Delft}
  \country{The Netherlands}
}

\author{Asterios Katsifodimos}
\email{a.katsifodimos@tudelft.nl}
\affiliation{%
  \institution{Delft University of Technology}
  \city{Delft}
  \country{The Netherlands}
}

\author{Rihan Hai}
\email{r.hai@tudelft.nl}
\affiliation{%
  \institution{Delft University of Technology}
  \city{Delft}
  \country{The Netherlands}
}

\renewcommand{\shortauthors}{Sun et al.}

\begin{abstract}
The rapid growth of large-scale machine learning (ML) models has led numerous commercial companies to utilize ML models for generating predictive results to help business decision-making. As two primary components in traditional predictive pipelines, data processing, and model predictions often operate in separate execution environments, leading to redundant engineering and computations. Additionally, the diverging mathematical foundations of data processing and machine learning hinder cross-optimizations by combining these two components, thereby overlooking potential opportunities to expedite predictive pipelines.

In this paper, we propose an operator fusing method based on GPU-accelerated linear algebraic evaluation of relational queries. Our method leverages linear algebra computation properties to merge operators in machine learning predictions and data processing, significantly accelerating predictive pipelines by up to 317x. We perform a complexity analysis to deliver quantitative insights into the advantages of operator fusion, considering various data and model dimensions. Furthermore, we extensively evaluate matrix multiplication query processing utilizing the widely-used Star Schema Benchmark. Through comprehensive evaluations, we demonstrate the effectiveness and potential of our approach in improving the efficiency of data processing and machine learning workloads on modern hardware.



\end{abstract}



\maketitle

\section{Introduction}

In recent years we are witnessing unprecedented growth in large-scale ML applications fueled by rapid advancements in computational capabilities, sophisticated models, and the increasing availability of vast amounts of data. Enterprises are now utilizing predictive results to assist in business decision-making and product design for customers. For instance, banks employ ML models for credit scoring and fraud detection, while online retailers use customers' historical behavior to provide real-time recommendations. In this thriving context, predictive ML applications call for efficient computation to meet the growing demands for real-time ML predictions and the substantial data processing workload required by ML models.

\para{Pitfalls of separating data processing and ML predictions}  Plenty of research efforts and commercial products have provided various solutions to accelerate data processing \cite{heavy,blazing} and ML predictions \cite{tvm,pytorch} using modern hardware like Graphics Processing Units (GPU). Thanks to massive parallelism and LA-friendly hardware architectures, the throughput of data processing and model predictive pipelines has significantly improved. However, the mixture of relational operators in data processing pipelines and Linear Algebraic operators in ML models introduces diverse data structures and software stacks. Specifically, data processing typically involves tasks such as data transformation and aggregation, which are traditionally solved with relational query engines. In contrast, model prediction workloads involve vast linear algebraic operations. The distinct mathematical foundations of data processing and model predictions often result in using separate software tools, libraries, and hardware configurations, which can hinder overall performance and efficiency. \textit{This separation can increase complexity, higher development and maintenance costs, and potential performance bottlenecks.}

\para{Mathematical gap of RA and LA} The different mathematical foundations present challenges for cross-optimizations when merging relational and linear algebra. Relational operators primarily process input data as sets of tuples, while LA computations operate on ordered scalars, vectors, and matrices. The data transformation and I/O cost between these two algebra systems result in significant overhead. Additionally, the diverging algebra systems imply different logical optimization strategies. Specifically, relational algebra (RA), a specification of first-order logic, can utilize logical reduction to decrease computational complexity. In contrast, LA operators can often take advantage of the numerical information of input matrices to reduce the size of intermediate results and overall complexity. \textit{In short, the   foundational differences between LA and RA obstruct further optimization by combining these two systems at both the logical and implementation levels.}

\para{RA operators on top of LA with GPU acceleration} A unified data representation and operators are desirable for ML practitioners. A promising new approach to address the challenges associated with the inconsistencies between LA and RA is to \emph{process relational data queries using linear algebra}  operations, such as matrix-matrix and matrix-vector multiplication. We term these queries Linear Algebra Queries (LAQ). By reformulating RA operations as LA operations, this approach can help bridge the gap between data processing and ML domains. Matrix multiplication is a linear algebra operation that can be efficiently parallelized and optimized using modern hardware, such as GPUs, which are designed to handle large-scale LA computations. By translating relational data queries into matrix multiplication operations, this approach can take advantage of the inherent parallelism and computational power of GPUs, leading to significant improvements in efficiency and scalability for both data processing and ML predictions.

Some operators (e.g., join, aggregation) have already been implemented and evaluated in recent studies \cite{joinproject1,joinproject2,joinproject3,TCU,wenbo_hard}. However, these studies do not compare their performance with full-fledged GPU databases, nor do they incorporate their methods into predictive pipelines involving machine learning predictions. As a result, \textit{the potential performance gains and practical implications of their methods in the end-to-end data processing and ML predictive pipelines remain unclear.}
\vspace{-1mm}
\begin{figure}[h!]
\centering
\includegraphics[width=2.8in]{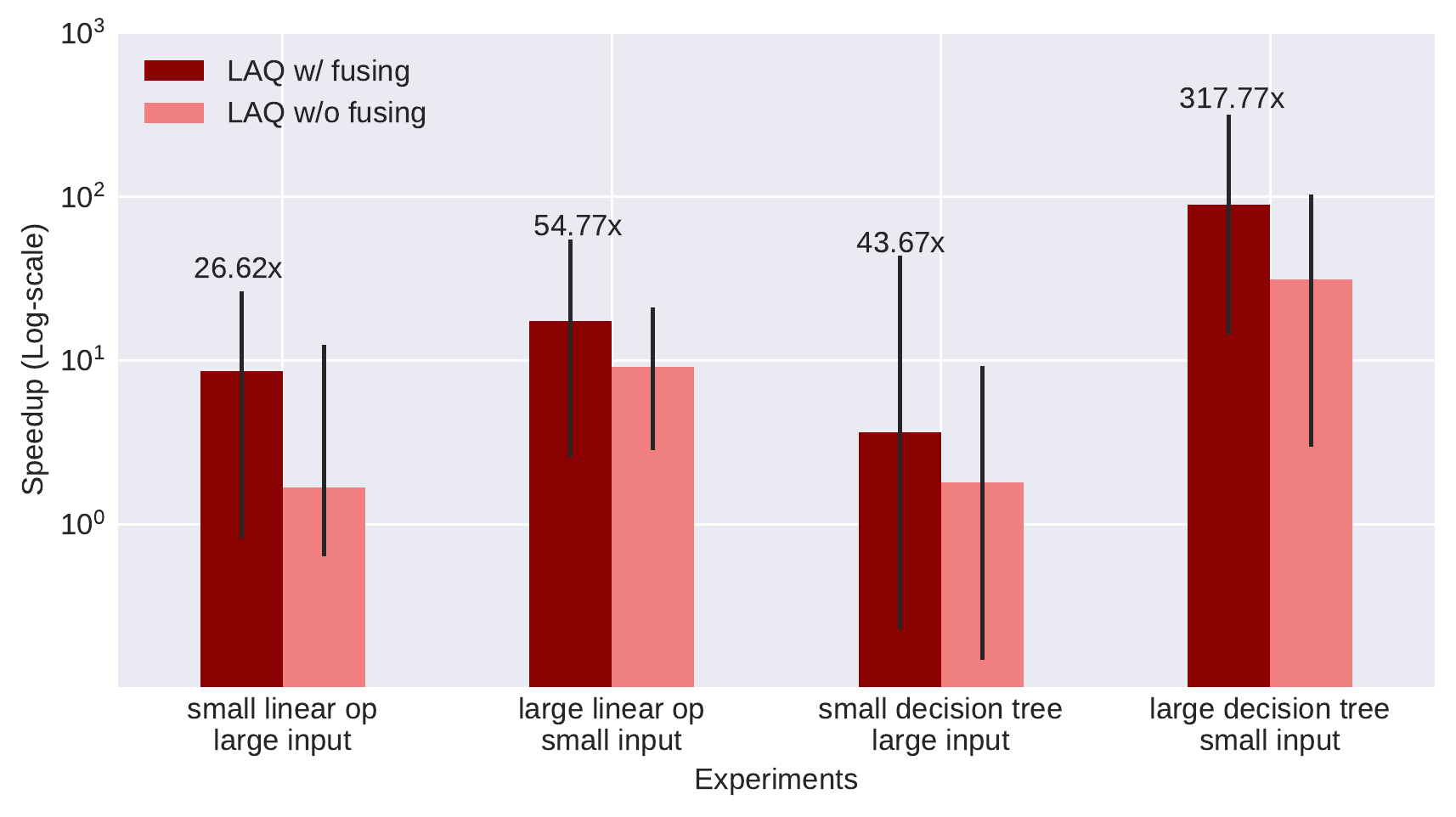}
\vspace{-1mm}
\caption{Speedups of our operator fusion method in four experimental predictive pipelines. The baseline is cuDF without operator fusion. The maximum attainable speedup is 317.77x.}
\label{fig:overview_speedup}
\end{figure}
\vspace{-1mm}

In this work, we propose a new approach to optimize performance of ML prediction following relational queries. The new approach leverages the unified representation in LAQ and fundamental properties of LA computations. The contributions of this work can be summarized as follows:
\vspace{-0.5mm}
\begin{itemize}
    \item We integrate batch model predictions into LAQ through operator fusion. By leveraging the computation properties of LA (i.e., associativity), we push down linear operators in ML models to dimension tables in a star schema \cite{kimball2011data} and merge operators in LAQ and models before prediction. Our operator fusion method achieves up to 317x speedups when evaluated on synthetic star schemas, as shown in Figure \ref{fig:overview_speedup}, compared to the separate execution of queries and predictions.

    \item We present a complexity analysis for operator fusion in the context of star schema queries followed by model predictions. This analysis provides quantitative insights into the benefits that can be gained from operator fusion, given specific data and model sizes.

    \item We thoroughly evaluate LAQ using the widely-adopted SSB data processing benchmark \cite{ssb} and report the performance comparison with cuDF~\cite{cudf} and HeavyDB~\cite{heavy}, two popular GPU-accelerated data processing systems. Our evaluation helps demonstrate the potential of LAQ in handling traditional data query workloads and its effectiveness in the context of end-to-end predictive pipelines.
\end{itemize}

This paper is organized as follows: Section 2 introduces primary operators in LAQ. These building blocks are based on existing works \cite{TCU,hai2023amalur}. Following that, Section 3 presents two examples to demonstrate the usefulness of our approach in predictive pipelines, which integrates linear operators in ML models into LAQ based on computation properties of LA. In Section 4, we first evaluate the performance of LAQ using the SSB dataset, and then we test the efficiency of our operator fusion method with synthetic datasets and models. In the final section, we provide insights into our research findings through experiments and discuss potential research directions derived from this study.

\section{Preliminaries: Linear algebra based query processing}
\label{LAQ}
This section introduces the approach of processing relational queries with linear algebra (LAQ). As the preliminaries to our operator fusion method in Section~\ref{fusing_method}, we implement the LAQ based on existing solutions. In particular, Section \ref{projection} and \ref{selection} elaborates selection and projection operator proposed in our earlier work \cite{hai2023amalur}. Section \ref{equijoin} -  \ref{sorting} introduces MMJoin, group-by aggregation and sorting operators in TCUDB \cite{TCU} and TQP \cite{tqp}. 

For clarity, we refer to the input tables of an RA operator (e.g., projection, join) as \emph{source tables} and the query results after executing the relational algebras as \emph{target tables}. Before we perform LAQ, all input tables are transformed into matrices for subsequent LA operators.



\vspace{0.2cm}
\begin{table}[h]
\caption{Important notations in Section 2 and 3}
\label{tab:notations}
\footnotesize
\begin{tabular}{|l|l|}
\hline
\textbf{Notations}      & \textbf{Description}                                 \\ \hline
$c$             & \#columns of a table                        \\ \hline
$r$             & \#rows of a table                           \\ \hline
$i$             & \#rows of the target table after joining    \\ \hline
$k$             & \#columns of the target table after joining \\ \hline
$p$             & \#features of a decision tree               \\ \hline
$l$             & \# output shape of models                \\ \hline
$\mathbf{v}$             & feature predicates of a decision tree       \\ \hline
$\mathbf{h}$             & values of leaves in a decision tree         \\ \hline
$M$             & schema mapping matrix                       \\ \hline
$I$             & row mapping matrix                          \\ \hline
$L$             & a simple linear operator                    \\ \hline
$F$             & feature mapping matrix of decision tree     \\ \hline
$H$             & paths to leaves in a decision tree          \\ \hline
$T$             & target table after joining                  \\ \hline
$R$, $S$, $B$, $C$, $D$ & tables                                      \\ \hline
\end{tabular}
\end{table}
\vspace{-0.2cm}
\subsection{Projection}
\label{projection}
We can effectively address the projection operator
using matrix multiplication. Projection entails extracting multiple columns from the source table and obtaining the target table. We compute projection through matrix multiplication by defining a \emph{column mapping matrix} $M \in \{0,1\}^{c \times k}$ \cite{hai2023amalur}, where $c$ is the number of columns of the source table and $k$ denotes the number of projected columns. As a preparation step, we add ID numbers to columns in the source and target tables. We define $M$ as follows:

\small{
\begin{align*}
    \begin{split} 
    M[i,j] =  \begin{cases}1, &if\ j^{th}\ column\  is\ the\ i^{th}\ column\  after\  selection\\
    0, & otherwise\end{cases}
    \end{split}
\end{align*}
}

\begin{figure}[h!]
\centering
\includegraphics[width=0.8\linewidth]{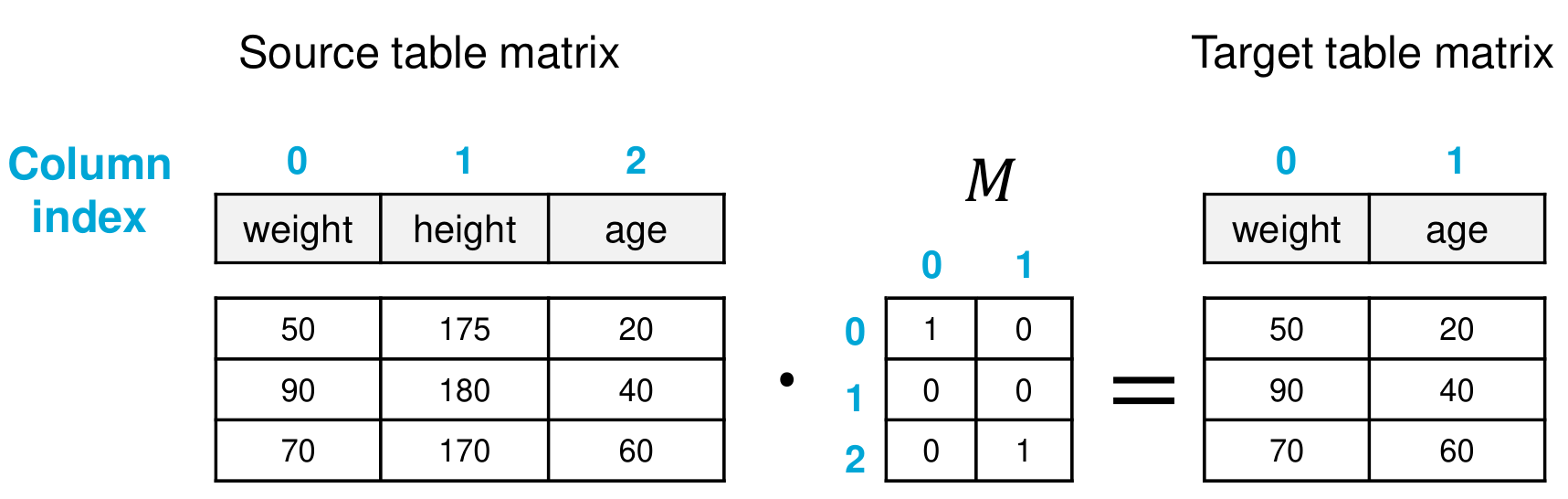}
\caption{An example of projection as matrix multiplication.}
\label{fig:colmap}
\end{figure}

Within one matrix $M$, for each projected column, its location in the target table after projection is represented by the vertical index $i$, while its original location in the source table is denoted by the horizontal index $j$. Non-zero values within the matrix $M$ indicate column correspondences between the source and target tables. As the source table has been converted to a matrix, column projection can be evaluated by multiplying the source table matrix and $M$.
Figure \ref{fig:colmap} shows an example of the projection process: given source table $Patient(weight, height, age)$, the projection operator $\pi _{weight, age}(Patient)$ is transformed to the matrix multiplication of source table matrix and column mapping matrix $M$. $M$ indicates that the columns with indexes of 0 and 2 of a source table are mapped to columns 0 and 1 of the target table.

\subsection{Selection}
\label{selection}
The selection operator produces a subset of tuples based on specific conditions, essentially filtering rows of the source table according to these conditions. We use a binary vector with the same length as the number of rows in the source table matrix to achieve the function of selection with linear algebra. For each row, the corresponding entry in the selection vector will be 1 if the row satisfies the selection condition and 0 otherwise. Multiplying the source table matrix with the selection vector (or its transpose, depending on the orientation of the matrices) effectively filters out rows that do not meet the selection criteria. The resulting matrix will contain only the rows that satisfy the selection condition.

\para{Improvement and implementation} However, this method may require an additional pass of column scan to generate the filter vector in advance, potentially making it less efficient than traditional relational selection. Thus, in our implementation, the multiplication with the filter vector is achieved using vectorized predicate 'AND' and memory copy rather than floating-point operations. In particular, if the input matrix has a row-dominated layout in memory, we select target row-pointers according to the filter vector and copy the selected rows to the target memory space. In our implementation, we use an out-of-box mask\_select operator provided by CuPy \cite{cupy}.

\subsection{MM-Join}
\label{equijoin}
\begin{algorithm}[h]
\caption{Matrix Multiplication Join}
\label{alg:mmjoin}

\SetKwInOut{Input}{Input}
\SetKwInOut{Output}{Output}
\SetKwComment{Comment}{// }{}
\Input{
$R$,$S$: input relations
}
\Output{
$I$: sparse matrix indicating maping rows in $R$ and $S$ 
}

key\_domain=union($key_R$, $key_S$)  //CUDA reduce

key\_len=len(key\_domain)

key\_dict = dict(zip(key\_domain range(0, key\_len))

$rows_R$ = range(0, $r_R$), $rows_S$ = range(0, $r_S$)

$columns_R$ = 0, $values_R$ = 1

$columns_S$ = 0, $values_S$ = 1

\For{i $\in [0,r_R)$}{
$columns_R$[i] = key\_dict[$key_R$[i]]  //CUDA parallel
}

\For{i $\in [0,r_S)$}{
$columns_S$[i] = key\_dict[$key_S$[i]] //CUDA parallel
}

$MAT_R$ = cuda\_construct\_CSR($rows_R$, $columns_R$, $values_R$)

$MAT_S$ = cuda\_construct\_CSR($columns_S$, $rows_S$, $values_S$)

$I$ = cuda\_sparse\_multiplication($MAT_R$, $MAT_S$).to\_COO()

\Return $I$
\end{algorithm}

The Matrix Multiplication Join (MM-Join) method takes advantage of matrix multiplication to evaluate join operations, which can be particularly beneficial when working with large datasets or when using hardware optimized for matrix multiplication, such as GPUs. This section introduces the MM-Join implementation in TCUDB \cite{TCU}. Apart from the implementation details, we discuss the computational complexity of the MM-Join and hash joins. To ensure portability and compatibility with machine learning workloads, we implement this algorithm using CuPy.

\subsubsection{2-way join}
\begin{figure}[t!]
\centering
\includegraphics[width=2in]{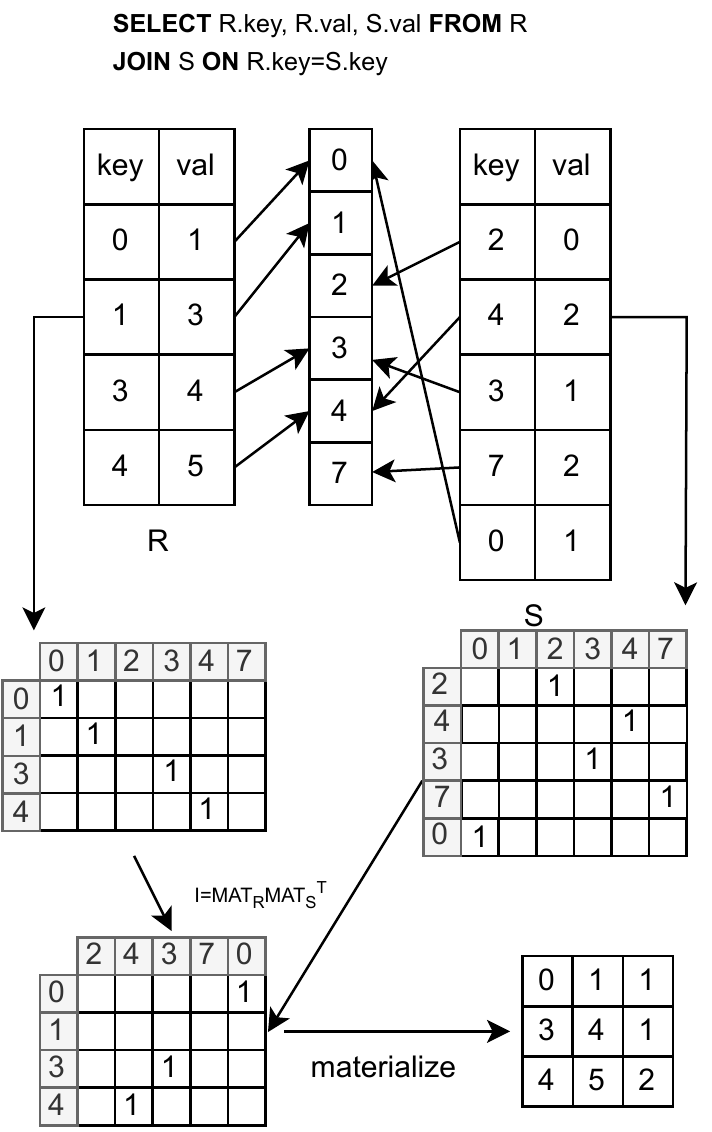}
\caption{An illustration for evaluating equi-join with matrix multiplication.}
\label{fig:mmuj}
\end{figure}

We illustrate the process of MM-Join with the pseudo-code in Algorithm \ref{alg:mmjoin}, which has four steps. We explain Algorithm \ref{alg:mmjoin} with the running example in Fig. \ref{fig:mmuj}. 
\\ $1)$ Suppose $R$ and $S$ are two tables to be joined, we first calculate the maximum key value in $R$ and $S$ to construct the common domain (Lines 1-3), resulting in $\{0, 1, 2, 3, 4, 7\}$; 
\\$2)$ Then we fill non-zero values and positions in sparse matrix format\footnote{We implement the sparse matrices in SciPy CSR: \url{https://docs.scipy.org/doc/scipy/reference/generated/scipy.sparse.csr_matrix.html}.} to get  $MAT_R$ and $MAT_S$ (Lines 4-12), which are sparse matrices storing relationships between keys and the common domain. The column indexes of the matrices are identical to the keys' positions in the common domain, and the row indexes are the row numbers of keys in original relations; 
\\$3)$ We execute sparse matrix multiplication over $MAT_R$ and transposed $MAT_S$ (Line 13); 
\\$4)$ The result $I$ is a row matching matrix\footnote{Implemented in SciPy COO format: \url{https://docs.scipy.org/doc/scipy/reference/generated/scipy.sparse.coo_matrix.html}}, defined as follows. 
\small{
\begin{align*}
    \begin{split} 
    I[i,j] =  \begin{cases}1, &if\ i^{th}\ row\ of\ R\ matches\ the\ j^{th}\ row\ of\ S\\
    0, & otherwise\end{cases}
    \end{split}
\end{align*}
}

The row-column pairs with non-zero values are matched rows in $R$ and $S$.

The high computational complexity and memory consumption have hindered the application of MM-Joins in CPU-based databases. In Algorithm \ref{alg:mmjoin}, transforming relations to matrices requires extra time and memory space based on the number of tuples and distinct keys, which is infeasible for relations with a large number of rows. 

\para{Approach analysis} The domain generation and retrieving process required by constructing sparse matrices involves a set union and two binary search in a sorted array, leading to a computational complexity as  $O(n^2\log n)$. Moreover, even though the CSR format can reduce  memory usage, the computational complexity of sparse matrix multiplication (spMM) can not be further reduced: the best-known complexity of spMM is $O(n^2)$\footnote{The complexity of spMM depends on matrix shapes and sparsity. Here we use an approximate value to show the complexity gap between spMM and hash join.} \cite{sparse}, which is higher than $O((|R|+|S|)*log(|R|))$ of a radix hash join algorithm \cite{parallelradix}, where $|R|$ and $|S|$ represent the cardinalities of the two tables participating the join.

Nevertheless, MM-Joins present an optimization opportunity that allows for the integration of linear operators in ML models with join processing. Conventionally, the results of relational queries need to be materialized before being utilized in model predictions. However, due to the LA representation of relational join processing, we can leverage LA optimization techniques, such as multiplication re-ordering, to reduce computational complexity and memory usage associated with redundant materialization. This integration can potentially improve the overall efficiency of combining relational operations with models.
\vspace{-0.1cm}
\subsubsection{Multi-way join}
In principle, multi-way joins can be naturally extended from 2-way joins through iterative evaluation following a given order. However, this naive implementation involves the materialization of intermediate tables, overlooking potential optimization opportunities hidden in the selectivity of join operators. In contrast, we can skip the materialization and use the matrix $I$ to evaluate subsequent joins. For instance, let's assume a join order of $Q$, $R$, and $S$. The matching rows of $Q$ and $R$ are stored in matrix $I_{QR}$. The rows that fail to match $Q$ will not appear in the final result. Therefore, we can directly use the matching row IDs of $R$ to join with $S$ and generate the matching matrix $I_{RS}$. This approach can enhance the performance between $R$ and $S$ due to the potential low selectivity of $Q \bowtie R$.
\vspace{-0.1cm}
\subsubsection{Materialization}
Now we use the  row matching matrix $I$ to preserve the matching row IDs of intermediate join results. We could use the IDs to generate a binary vector and treat the materialization as a selection using the method described in Section \ref{selection}. 

However, this non-LA operation hinders further optimizations by integrating ML models with joins. Alternatively, a materialized table can be viewed as a result of the projection of transposed source tables. In this regard, we need to construct the mapping matrix $M$ using the result matrix $I$ from MM-Join. Consequently, we require two row sparse mapping matrices for relations $R$ and $S$, as follows:

\small{
\begin{align*}
\label{eq:I}
    \begin{split} 
    I_*[i,j] =  \begin{cases}1, &if\ j^{th}\ row\ of\ R\ is\ the\ i^{th}\ row\ of\ materialized\ table\\
    0, & otherwise\end{cases}
    \end{split}
\end{align*}
}

The COO format of $I$ has three attributes: row indexes, column indexes, and the number of non-zero values ($nnz$). The $nnz$ is precisely the number of rows in the materialized table, which implies that the row IDs of the materialized table can be represented as a vector $m = <0, 1, 2, \dots, nnz - 1>$. Consequently, we can construct two row mapping matrices $I_R$ and $I_S$ by aligning row indexes and column indexes with $m$, respectively.
\vspace{-0.2cm}
\subsection{Group-by Aggregation}
In certain analytical queries, we may need to perform aggregation based on specific values after a join operation. Employing a similar technique used for join processing, we can also represent group-by operations concerning a single attribute through matrix multiplication. However, we cannot directly compute multi-aggregation following multi-way joins because it lacks the capability to express value interaction among attributes. In this section, we explain the single-column aggregation in TCUDB \cite{TCU} and the multi-column aggregation inspired by TQP \cite{tqp}. 
\vspace{-0.1cm}
\subsubsection{Aggregation by a single column}
\begin{figure}[t!]
\centering
\includegraphics[width=2in]{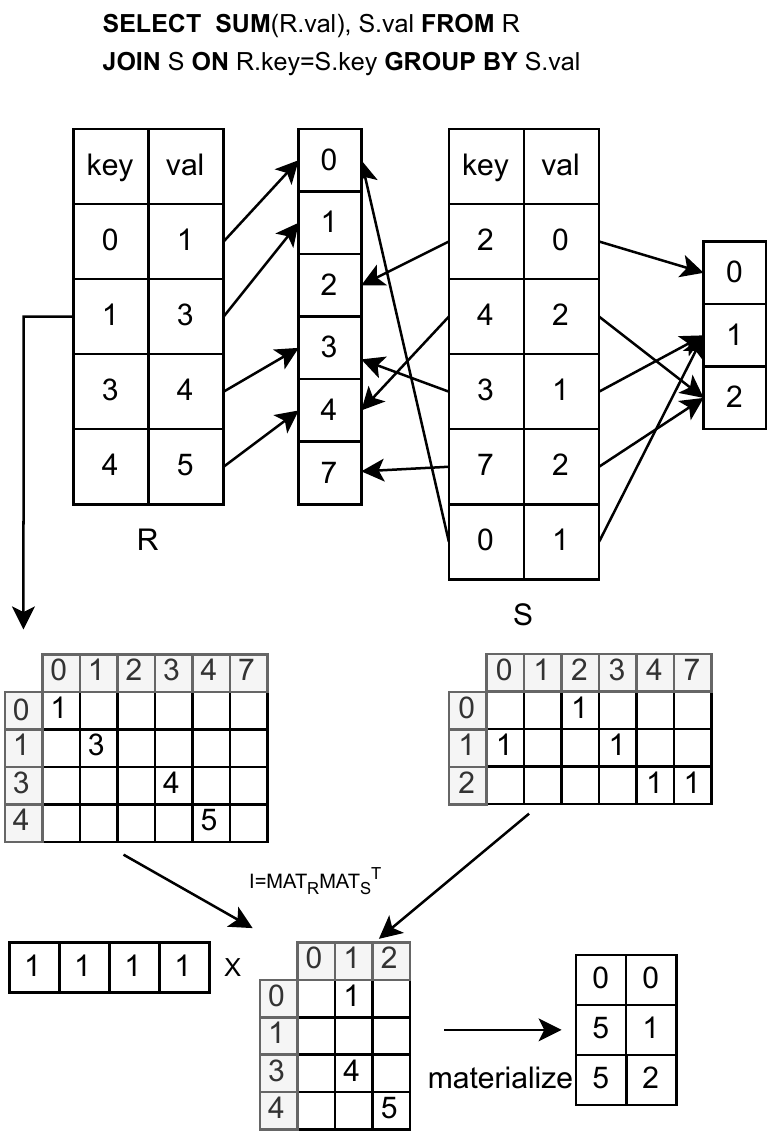}
\caption{An illustration for evaluating group-by-sum with matrix multiplication. }
\label{fig:mmagg}
\end{figure}
Figure \ref{fig:mmagg} demonstrates how to evaluate single attribute aggregation using LA. The fundamental pattern is similar to the MM-Join, but two sparse matrices require some adjustments. First, $MAT_R$ is no longer a binary matrix; the value column of $R$ to be aggregated is filled into the sparse matrix $MAT_R$. 

As for table $S$, we begin by finding  unique values as groups. $MAT_S$ is filled with values of 1, according to relationships between groups and the key domain. In the example presented in Figure \ref{fig:mmagg}, values 0, 1, 2 are found as groups. Then we find relationships between keys of $S$ and the groups, which are $\{2\}->0$, $\{3,0\}->1$, $\{4,7\}->2$. After filling 1s according to the relationships, relationships between keys in $R$ to the group can be evaluated by multiplying $MAT_R$ and ${MAT_S}^T$. Finally, to perform summation of values in $R$, we introduce a reduction vector filled with values of 1, enabling the materialization of the result.
\vspace{-0.1cm}
\subsubsection{Aggregation by multiple columns}
Aggregation by multiple columns cannot be directly integrated with MM-Join in the same way as single-column aggregation. As shown in Figure \ref{fig:mmagg}, we require a matrix representing relationships between the key domain and groups. For single-column aggregation, groups can be evaluated using a numerical unique function. However, for multi-column aggregation, we must first join tables involved in groups and then apply the unique function to tuples, which is not consistent with other numerical operators.

To complete the queries for evaluation in our experiments, we adopt an alternative solution proposed in \cite{tqp}, where unique tuples are identified using a sort-unique procedure.
 \vspace{-0.2cm}
\subsection{Sorting}
\label{sorting}
Sorting cannot be directly represented in LA, but we can integrate sorting into MM-Join if the sorting is performed on keys to be joined. The column indices in matrices $MAT_S$ and $MAT_R$ correspond to the positions of keys in the key domain. As a result, by sorting the key domain, we can obtain $MAT_R$ and $MAT_S$ with sorted keys. This approach allows us to seamlessly integrate the sorting operation into the MM-Join process.

\para{Summary} This section discusses existing methods for evaluating relational operators using LAQ and identifies their limitations. Some operators, such as selection, projection, equi-join, and single-column aggregation, can be equivalently represented by linear algebraic computations. However, multi-column aggregation and sorting cannot be transformed into linear algebra operations. To address these limitations, we implement alternative GPU-compatible methods for these two operators, enabling LAQ to evaluate a wider range of relational queries. This allows us to explore the theoretical unification between data processing and downstream ML model predictions on GPUs.

\vspace{-0.1cm}
\section{Operator Fusion}
\label{fusing_method}
On the basis of LAQ, in this section, we propose an operator fusion method to merge operators in ML model predictions and LAQ for the speedup of predictive pipelines. Specifically, given the fact that operators in LAQ and ML predictions are uniformly represented as linear algebraic computations, we can leverage the computation properties of LA, such as associativity of matrix multiplication, to reduce computational complexity or the size of intermediate results. Moreover, by utilizing the distributive property of matrix multiplication, ML operators can be pushed down to source tables and stored as matrices, subsequently decreasing the computational complexity of real-time predictions.

In this section, based on the LA operators introduced in Section~\ref{LAQ}, we analyze two operator fusion with two ML examples to illustrate the benefits: fusing linear operators (Section~\ref{case_linear}), and decision trees (Section~\ref{case_tree}). 
\vspace{-0.2cm}
\subsection{Scenario Description}
Given a data warehouse containing a star schema with a central fact table $A$ and dimension tables $B$, $C$, and $D$, we consider the following scenario. Fact table $A$ stores transactional data, while dimension tables $B$, $C$, and $D$ contain contextual attributes associated with the facts in table $A$. A star join operation is applied to join the fact table $A$ with dimension tables $B$, $C$, and $D$, leveraging their respective foreign key-primary key relationships. The resulting dataset $S$ from this star join operation integrates facts and dimension attributes. Subsequent ML operators take dataset $S$ as input to produce matrices for further applications.

\para{Operator Fusion} According to the design principles of star schema data warehouses discussed in \cite{kimball2011data}, fact tables tend to exhibit higher update frequencies and larger data volumes compared to dimension tables. Consequently, fusing downstream ML operators with relatively static dimension tables allows for pre-fusing of partial results, thereby reducing the cost of predictions. We term this approach  \emph{operator fusion}. In the following sections, we leverage two models to demonstrate how operator fusion can accelerate predictive pipelines.

\begin{figure*}[t!]
\centering
\includegraphics[width=0.83\textwidth]{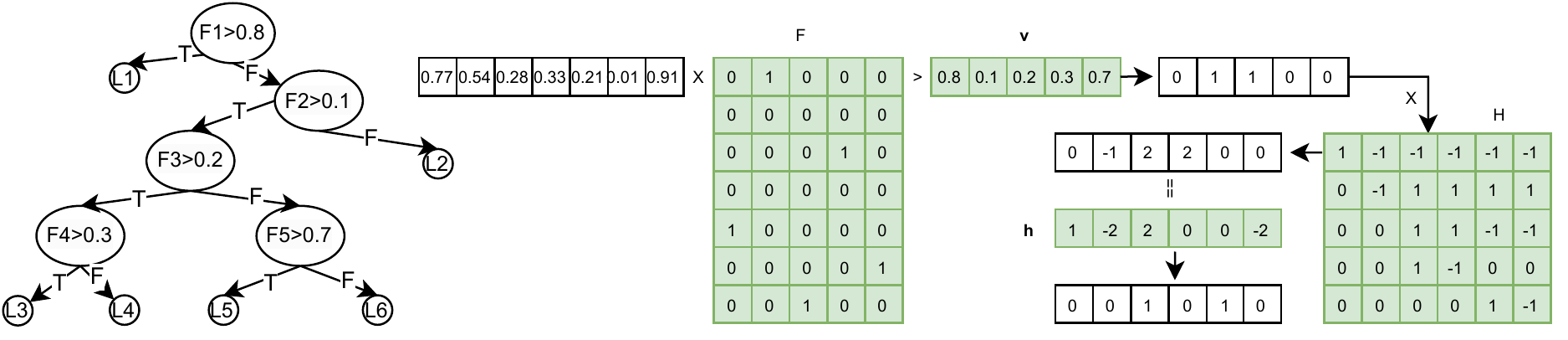}
\caption{Prediction with a decision tree with linear algebraic representation adapted from \cite{hummingbird1}.}
\label{fig:tree}
\end{figure*}
\vspace{-0.2cm}
\subsection{Fusing Simple Linear Operators}
\label{case_linear}
Suppose the result of a star join $T \in \mathbb{R}^{i \times k}$ is computed through MM-Join according to keys in the fact table $A$. The evaluation can be presented as $T=I_1BM_1+I_2CM_2+I_3DM_3$, where $I_1 \in \{0,1\}^{i\times r_1}$, $I_2 \in \{0,1\}^{i\times r_2}$, $I_3 \in \{0,1\}^{i\times r_3}$, and $M_* \in \{0,1\}^{c\times k}$.  $c$ and $k$ denote the number of columns in dimension tables and selected features for linear operators respectively. 

The result $S$ is then multiplied by a linear operator $L \in \mathbb{R}^{k \times l}$, resulting in predictions as $\mathbb{R}^{i\times l}$. For simplification, in the following analysis, we equally separate features into each dimension table, which means $c=\frac{k}{3}$.  According to the s associativity of  matrix multiplication, we can fuse $L$ to dimension tables using:
\begin{equation}
\label{eq:linear_op_origin}
\begin{aligned}
    \text{predictions} &= TL \\
    &=(I_1BM_1+I_2CM_2+I_3DM_3)L \\
    &=I_1\underline{BM_1L}+I_2\underline{CM_2L}+I_3\underline{DM_3L}
\end{aligned}
\end{equation}

We follow the common assumption that dimension tables are less frequently updated than the fact table, $BM_1L$, $CM_2L$ and $DM_3L$ can be treated as constants in a period. Therefore, we can \emph{pre-fuse} them and only apply row matching matrix $I_*$ when materialization.
\subsubsection{Complexity analysis} We now perform a complexity analysis for operator fusion in a predictive pipeline and compare it with non-fused methods. 
As both fused and non-fused methods share the same domain generation step, we will omit the complexity of domain generation in the following analysis for comparison purposes. Additionally, matrix additions have lower complexity order than matrix multiplications. Therefore, in the complexity analysis for this section and Section \ref{case_tree}, \emph{we omit the complexity of matrix additions}. Given the aforementioned dimensions of matrices, the computational complexity without operator fusion is:
\begin{align*}
    \mathcal{C}_{no-fusion} &=\mathcal{C}_{mmjoin} + \mathcal{C}_{op} \\
    &=ck\sum_j{r_j}+ik\sum_j{r_j}+ikl\\
    &=(ik+\frac{k^2}{3})\sum_j{r_j}+ikl
\end{align*}
If $L$ is pushed down to dimension tables, we will get three pre-fused partial values of the final result, $BM_1L \in \mathbb{R}^{r_1\times l }$, $CM_2L\in \mathbb{R}^{r_2\times l }$ and $DM_3L\in \mathbb{R}^{r_3\times l }$. The linear operator can be directly applied to these partial results. Then we have the computation complexity as:
\begin{align*}
        \mathcal{C}_{fusion} = il\sum_j{r_j}
\end{align*}
Now, we compare two complexity values:

\begin{equation}
\label{linear_complexity}
    \begin{aligned}
    \frac{\mathcal{C}_{non-fusion}}{\mathcal{C}_{fusion}}&=\frac{(ik+\frac{k^2}{3})\sum_j{r_j}+ikl}{il\sum_j{r_j}}\\
    &=\frac{k}{l}+\frac{k^2}{3il}+\frac{k}{\sum_j{r_j}}
    \end{aligned}
\end{equation}

Upon analyzing the information above, it becomes evident that the speedup of operator fusion is correlated with the shape of the linear operator and the cardinality of dimension tables. In practical predictive tasks, the total number of rows of dimension tables is usually much larger than the number of columns. Therefore, we can safely ignore the terms $\frac{k^2}{3il}$ and$\frac{k}{\sum_j{r_j}}$. In particular, $\frac{k}{l}$ can be considered as the filtering effect of the linear operator. For instance, a linear regression model can be viewed as a linear operator with an output shape of 1. By pre-fusing the linear regression model with dimension tables, the partial values to be composed after a join operation are vectors instead of matrices. Consequently, the execution time of the join-prediction operation can be significantly reduced. In Section \ref{exp_fusing}, we will examine the speedups of the fusion method concerning various input settings.

\vspace{-0.2cm}
\subsection{Fusing Decision Trees}
\label{case_tree}
Tree models, such as decision trees, are popular among data scientists due to their interpretability \cite{tree_pop}. In this section, we elaborate on our operator fusion method with more complex decision tree models and explore optimization opportunities using a matrix representation of decision tree predictions. For clarity, our method is built on the linear algebraic representation of decision trees proposed by Hummingbird \cite{hummingbird1} in Section \ref{humming_tree}.
\vspace{-0.5cm}
\subsubsection{Matrix representation of Descion Trees}
\label{humming_tree}
Hummingbird \cite{hummingbird1} introduces a method to represent decision tree models using linear algebra operators. The key idea is to transform the tree structure into a set of linear algebraic operations and vectorized predicates, which can then be efficiently executed on hardware optimized for such computations, like GPUs.

Suppose we have a batch of vectors $S \in \mathbb{R}^{i \times k}$. To represent a tree, we need two binary matrices, $F \in \{0,1\}^{k\times p}$ and $H\in \{-1,0,1\}^{f\times l}$, as well as two vectors, $\mathbf{f} \in \mathbb{R}^p$ and $\mathbf{h} \in \mathbb{R}^l$. Figure \ref{fig:tree} illustrates how to use a sequence of linear and predicate operators to perform a prediction for decision trees. The final result is a binary encoding for the prediction label, which can be subsequently retrieved through a lookup table. 


\para{Step 1}
The binary matrix $F$ is an orthogonal matrix that maps input vectors to features. Some columns may not be in the selected features; thus, the matrix serves as a feature selection operator. As the initial linear operator of the decision tree, its orthogonality enables operator fusion because the result of $TF$ is a linear combination of the original columns in the input.
In practice, $F$ can be integrated into the column mapping matrix $M_*$ (described in Section \ref{projection}), but we retain it in the rest of the implementation for completeness.

\para{Step 2} Vector $\mathbf{v}$ represents the values of nodes in the decision tree. The order of this vector follows a pre-assigned rank. The output of operator $F$ undergoes a predicate '$> \mathbf{v}$', producing a binary vector. In practice, we often apply predictions to a batch of vectors; as such, the output of this step turns out to be a matrix.
Notably, since the output of the last step is a linear combination of the original input, each column can be independently compared to the corresponding values in $\mathbf{v}$. This means that the predicate can be fused with dimension tables as well.

\para{Step 3} Matrix $H$ signifies the structure of decision trees. Each column represents the path of a leaf node. Values in the column indicate the choices of nodes on this path, where 1 means True, and -1 means False. For instance, the path to L2 contains two nodes, F1 and F2, both of which choose the False side. Consequently, the values of the corresponding nodes are -1.
Notably, $H$ is a reduction operator. Fusing $H$ with dimension tables is applicable but only produces a local sum.

\para{Step 4} Vector $\mathbf{h}$ is the column sum of matrix $H$. Before comparing with $\mathbf{h}$, the pre-fused matrices must be added to obtain a complete vector. The result vector is compared with $\mathbf{h}$, and a binary encoding denoting prediction labels will be produced.
\vspace{-0.1cm}
\subsubsection{Fusing with dimension tables}
As we discussed in last section, the prediction results of the decision tree over $T$ can be represented by:
\[
    predictions= \underbrace{\overbrace{((\underbrace{\overbrace{TF}^\text{step 1}>\mathbf{v}}_\text{step 2})H)}^\text{step 3}==\mathbf{h}}_\text{step 4}
\]
Because vectors in $F$ are orthogonal, the result of $TF$ can be interpreted as a linear combination of vectors in $T$. As a consequence, the predicate operator '$>\mathbf{v}$' can be partially evaluated. In contrast, the predicate operator '$==\mathbf{h}$' depends on the predecessor reduction operator $H$, which can not be partially evaluated. Therefore, we can push down $(TF>\mathbf{v})H$ to dimension tables and evaluate the predicate equal operator by summing up partial results. The process is expressed as follows:

\begin{equation}
\label{fusing_tree}
    \begin{aligned}
        predictions &= ((TF>\mathbf{v})H)==\mathbf{h}\\
    &=(I_1((BM_1F>\mathbf{v})H)  \\
    & \quad + I_2((CM_2F>\mathbf{v})H) \\
    & \quad +I_3((DM_3F>\mathbf{v})H))==\mathbf{h}\\
    &=(I_1T_1+I_2T_2+I_3T_3)==\mathbf{h}
    \end{aligned}
\end{equation}

\subsubsection{Complexity Analysis}
Similar to the complexity analysis for the linear operator, we first present the complexity of non-fusion method:
\begin{align*}
    \mathcal{C}_{non-fusion} &=\mathcal{C}_{mmjoin} + \mathcal{C}_{F}+\mathcal{C}_{\mathbf{v}}+\mathcal{C}_{H}+\mathcal{C}_{\mathbf{h}} \\
    &=(\frac{k^2}{3}+ik)\sum_j{r_j}+ikp+ip+ipl+il
\end{align*}

With operator fusion presented in equation \ref{fusing_tree}, we have three pre-fused matrices whose dimensions are $R^{r_i\times l}$. The complexity of remaining operations of decision tree's result can be expressed by:
\begin{align*}
    \mathcal{C}_{fusion}=il\sum_j{r_j}+il
\end{align*}
Then we compare the complexities through $\frac{\mathcal{C}_{no-fusion}}{\mathcal{C}_{fusion}}$, which is:
\begin{equation*}
    \frac{(\frac{k^2}{3}+ik)\sum_j{r_j}+ikp+ip+ipl}{(il+1)\sum_j{r_j}}
\end{equation*}
$r_j$ represents the number of rows in a dimension table, while $p$ denotes the number of features. For simplicity, we assume the number of features ($p$) equals to length of input ($k$). Additionally, considering $il>>1$, we remove the constant term. Then, we have:
\begin{equation}
\label{tree_comp_frac}
\begin{aligned}
   \frac{\mathcal{C}_{non-fusion}}{\mathcal{C}_{fusion}}
    &= \frac{\frac{k^2}{3}}{il}+\frac{ik}{il}+\frac{ik}{il\sum_j{r_j}}+\frac{ikl}{il\sum_j{r_j}}+\frac{1}{\sum_j{r_j}}\\
    &=\frac{k}{l}+\frac{k^2}{3il}+\frac{k^2}{l\sum_j{r_j}}+\frac{k}{\sum_j{r_j}}+\frac{k}{l\sum_j{r_j}}+\frac{1}{\sum_j{r_j}}
\end{aligned}
\end{equation}

Due to the involvement of more linear operators in decision trees, additional tail terms appear in Equation \ref{tree_comp_frac}. When the number of rows in dimension tables is smaller than the number of features, we can expect that our approach facilitates a certain speedup through operator fusion. In contrast, when the size of dimension tables is significantly larger than $k$, the speedup is correlated with the first term $\frac{k}{l}$. Similar to the discussion in Section \ref{case_tree}, the filtering effect of decision trees determines the benefit of operator fusion. If a tree has only a small number of leaves, a significant speedup can be expected. We will further substantiate this analysis with experimental results in Section~\ref{exp_fusing}.

Our analysis above assumes that all matrices involved in the computation are dense matrices. In our implementation, matrices $I$, $M$, and $F$ are stored in the CSR format and computed using a sparse matrix multiplication kernel. This approach has lower computational complexity compared to naive dense matrix multiplication, further enhancing the efficiency of the overall computation process.

\para{Summary} In this section, we proposed an operator fusion method to accelerate predictive pipelines, building on the preliminaries introduced in section \ref{LAQ}. Within the context of data warehouses, we presented two predictive pipelines that demonstrate how operator fusion accelerates them by pre-fused partial results. Additionally, we compare the theoretical complexity of fusion and non-fusion methods and identify a preliminary decision boundary for determining when to apply operator fusion for speedup. In Section \ref{exp_fusing}, we will present the speedup of operator fusion in predictive pipelines.


It is important to note that two examples in this section assume dimension tables are updated less frequently than the fact table, which is a common design principle in traditional data warehouses. However, many data architectures (e.g., data mesh \cite{datamesh}, data fabric \cite{datafab}) proposed in recent years have gradually deviated from this principle. Further investigation is needed to determine the applicability of the operator fusion method in these scenarios.


\section{Experimental Evaluation}
In this section, we evaluate the performance of LAQ and examine the performance enhancement achieved through operator fusion.  In Section \ref{exp_mmjoin}, we use the SSB \cite{ssb} dataset to compare the performance of LAQ with two GPU-accelerated relational query processing engines. Then, based on LAQ, we assess the speedup of operator fusion through two models introduced in Section \ref{case_linear} and \ref{case_tree}. Before presenting the experimental results, we first provide an overview of the experimental setup, encompassing the implementations under evaluation, dataset characteristics, and hardware.
\subsection{Experiment settings}

\para{Implementation}
In this paper, we implement GPU-accelerated LAQ using CuPy \cite{cupy}. The implementation involves two-way join, multi-way join, and bi-group aggregation, all of which are computed using CSR format with CuSparse, which is a CUDA library for sparse matrix multiplication\footnote{\url{https://docs.nvidia.com/cuda/cusparse/index.html}}. Although multi-group aggregation can be theoretically calculated through tensor multiplications, we opt to use the scatter\_add operator due to the absence of a suitable vendor library for sparse tensor multiplications.

\para{Baselines} To assess the performance of our LAQ implementation, we compare it with two other GPU-accelerated data processing libraries: HeavyDB \cite{heavy} and cuDF \cite{cudf}. HeavyDB, formerly known as OmniSciDB, is a commercial GPU data management system that supports a wide range of relational queries on GPUs. It features a query optimizer and cache strategy to expedite query execution. cuDF, on the other hand, is a GPU DataFrame library that provides support for commonly used relational operators, such as selection, projection, join, and aggregations. Unlike HeavyDB, cuDF is a vanilla query processor, similar to our LA query implementation, and serves as an appropriate baseline for our study.

\para{Workload} For the evaluation of operator fusion, we use star join queries on a synthetic dataset characterized in Table \ref{tab:synth_card}, and the results are subsequently utilized as input for linear operators. 

\para{Hardware} All the implementations in these experiments are executed on an Nvidia A40 (48GB) GPU, eliminating the need to account for the communication cost between host memory and device memory. Each experiment is conducted ten times, and we report the mean values and standard errors.
\vspace{-0.1cm}
\subsubsection{Datasets}
We use Star Schema Benchmark (SSB) \cite{ssb} to investigate the performance in real-world workloads. On the other hand, considering the memory space required by linear operators in the operator fusion evaluation, we generate a synthetic dataset with down-scaled cardinality of SSB dataset. In the following parts, we introduce these two datasets and their data characteristics.

\para{Star Schema Benchmark}
The SSB dataset is a widely-used benchmark for evaluating the performance of data warehouse systems and database management systems. It was developed as a simplified version of the TPC-H benchmark, which is also designed for testing data warehouse systems. The SSB dataset focuses on star schema query processing and comes with a predefined set of queries that test various aspects of database performance, such as join operations, aggregations, and filtering. 

\begin{figure}[h!]
\centering
\includegraphics[width=2.8in]{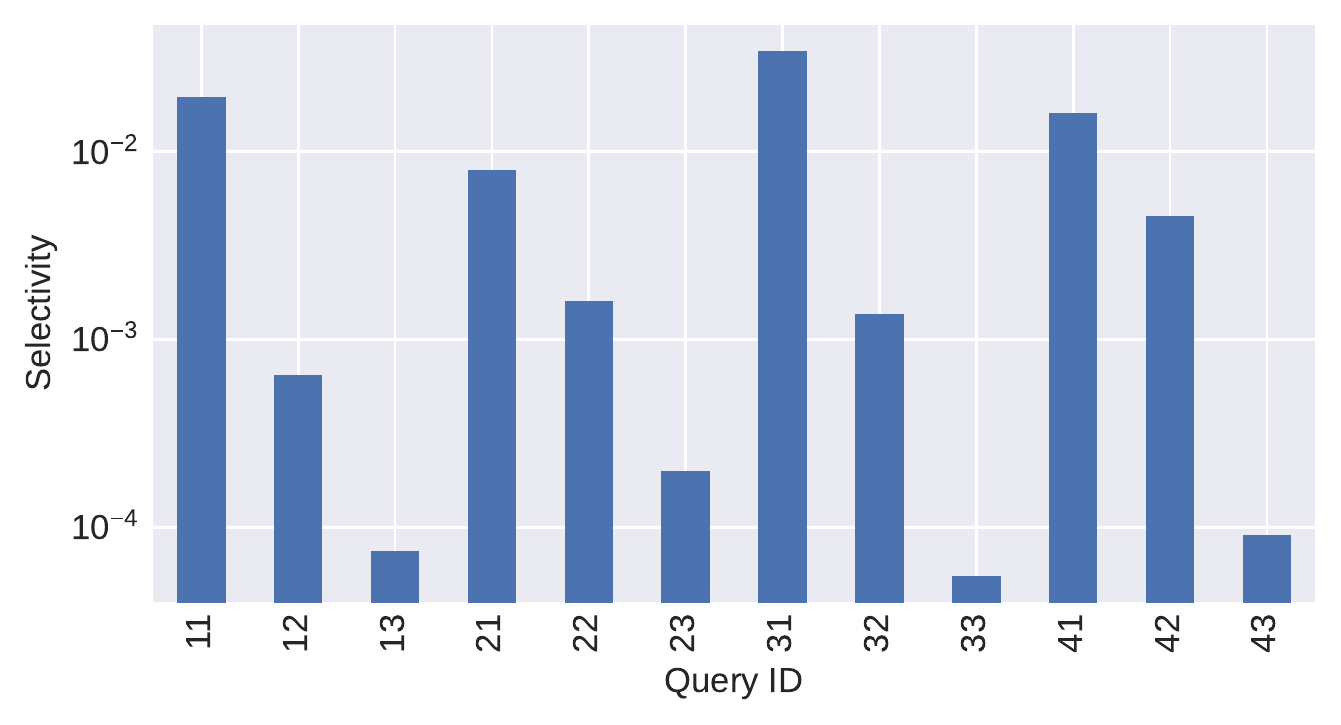}
\caption{Selectivity of each query in SSB}
\label{fig:sels}
\vspace{0.2cm}
\end{figure}

Table \ref{tab:SSB_quries} provides a summary of the workloads and query groups, while Table \ref{tab:SSB_card} displays the types and cardinality settings for each table in the SSB dataset. Additionally, Figure \ref{fig:sels} illustrates the selectivities of each query for subsequent evaluations. The parameter $sf$ represents the scale factor that controls data sizes, and it will be used to denote the scale of data throughout the rest of the paper.
\begin{table}[h!]
\caption{Summary for query groups in SSB}
\label{tab:SSB_quries}\
\footnotesize
\begin{tabular}{|l|l|l|l|l|}
\hline
\textbf{Queries}      & \textbf{Group 1} & \textbf{Group 2}                                                 & \textbf{Group3}                                                 & \textbf{Group 4}                                                 \\ \hline
ID of subqueries     & 11, 12, 13      & 21, 22, 23    & 31, 32, 33       & 41, 42, 43                                                      \\ \hline

\# Joins     & 1      & 3                                                      & 3                                                      & 4                                                      \\ \hline
Aggregations & Sum    & \begin{tabular}[c]{@{}l@{}}Group-by\\ Sum\end{tabular} & \begin{tabular}[c]{@{}l@{}}Group-by\\ Sum\end{tabular} & \begin{tabular}[c]{@{}l@{}}Group-by\\ Sum\end{tabular} \\ \hline
Sorting      & No     & Yes                                                    & Yes                                                    & Yes                                                    \\ \hline
\end{tabular}
\end{table}
\vspace{0.2cm}
\begin{table}[h!]
\caption{Types and cardinalities of SSB tables. $sf$ is a parameter controlling data sizes.}
\label{tab:SSB_card}
\footnotesize
\begin{tabular}{|l|l|l|}
\hline
\textbf{Tables}    & \textbf{Type} & \textbf{Cardinality}              \\ \hline
lineorder & Fact & $sf*6,000,000$             \\ \hline
part      & Dim  & $200,000*floor(1+\log_2 sf)$ \\ \hline
supplier  & Dim  & $sf*2,000$                 \\ \hline
customer  & Dim  & $sf*30,000$              \\ \hline
date      & Dim  & $7*365$                    \\ \hline
\end{tabular}
\end{table}

\para{Synthetic dataset}
SSB queries are well-suited for evaluating operator fusion, as our scenario setting aligns with the design principles of SSB. Because allocating SSB dataset and models to be evaluated in the same GPU causes out-of-memory error, we generate a synthetic dataset based on down-scaled cardinalities of SSB tables for operator fusion experiments. We introduce two groups of cardinality settings to test the performance of operator fusion with varying numbers of input rows. The detailed table design is shown in Table \ref{tab:synth_card}.
\begin{table}[h!]
\vspace{0.2cm}
\caption{Types and cardinalities of synthetic tables. sf is a parameter controlling data sizes.}
\label{tab:synth_card}
\footnotesize
\begin{tabular}{|l|l|l|l|}
\hline
\textbf{Tables}    & \textbf{Type} & \textbf{Cardinality setting 1} & \textbf{Cardinality setting 2}              \\ \hline
lineorder & Fact & $sf*600,000$  &   $sf*3,000$        \\ \hline
part      & Dim  & $20,000*floor(1+\log_2 sf)$ &$2,000*floor(1+\log_2 sf)$ \\ \hline
supplier  & Dim  & $sf*2,000$  & $sf*2,000$              \\ \hline
date      & Dim  & $7*365$  &$7*365$                    \\ \hline
\end{tabular}
\end{table}

In addition to varying settings in Table \ref{tab:synth_card}, we also alter the size of models to be fused in order to test the performance of operator fusion under different computing workloads. To match the input shape $k$ of models, we adjust the number of columns accordingly. The parameters of linear operators are demonstrated in Tables \ref{tab:param_case}.

\vspace{0.2cm}
\begin{table}[t!]
\caption{Parameters for linear operator and decision tree}
\label{tab:param_case}
\footnotesize
\begin{tabular}{|lllll|}
\hline
\multicolumn{5}{|c|}{\textbf{Simple Linear Operator}}                                                                                                                                                                                 \\ \hline
\multicolumn{1}{|l|}{\textbf{Cardinality}} & \multicolumn{1}{c|}{\textbf{sf}}               & \multicolumn{1}{l|}{\textbf{length of input} ($k$)}                        & \multicolumn{2}{l|}{\textbf{length of output} ($l$)}                                       \\ \hline
\multicolumn{1}{|l|}{\textbf{Setting 1}}     & \multicolumn{1}{l|}{{}1,2,4,8,16{}} & \multicolumn{1}{l|}{$2^{[4 \dots 7]}$}      & \multicolumn{2}{l|}{$2^{[1 \dots 7]}$ }   \\ \hline
\multicolumn{1}{|l|}{\textbf{Setting 2}}     & \multicolumn{1}{l|}{1,2}              & \multicolumn{1}{l|}{$2^{[8 \dots 11]}$} & \multicolumn{2}{l|}{$2^{[1 \dots k]}$} \\ \hline
\multicolumn{5}{|c|}{\textbf{Decision Tree}}                                                                                                                                                                                   \\ \hline
\multicolumn{1}{|l|}{\textbf{Cardinality}} & \multicolumn{1}{c|}{\textbf{sf}}               & \multicolumn{1}{l|}{\textbf{length of input} ($k$)}                        & \multicolumn{1}{l|}{\textbf{\# features} ($p$)}                 & \textbf{\# leaves} ($l$)                \\ \hline
\multicolumn{1}{|l|}{\textbf{Setting 1}}             & \multicolumn{1}{l|}{{}1,2,4,8,16{}}                & \multicolumn{1}{l|}{$2^{[4 \dots 7]}$}                                       & \multicolumn{1}{l|}{$2^{[4 \dots 7]}$}                            &         $2^{[1 \dots 6]}$                 \\ \hline
\multicolumn{1}{|l|}{\textbf{Setting 2}}             & \multicolumn{1}{l|}{1,2}                  & \multicolumn{1}{l|}{$2^{[8 \dots 11]}$}                                       & \multicolumn{1}{l|}{$2^{[3 \dots 11]}$ }                            &            $2^{[6 \dots 11]}$                \\ \hline
\end{tabular}
\end{table}

\subsection{Performance Evaluation for LAQ}
\label{exp_mmjoin}

In this section, we evaluate the performance of LAQ using the SSB dataset on GPUs and compare the results with two GPU-accelerated data processing engines. First, we measure the average execution time with respect to varying scale factors ($sf$), and then we examine LAQ's performance on different queries. To identify the most time-consuming operator, we provide a performance breakdown and suggest an optimization opportunity for future research.

\mybox{$Q_1:$ When does LAQ perform better than cuDF and HeavyDB w.r.t. varying data sizes?}
\begin{figure}[t!]
\centering
\includegraphics[width=2.8in]{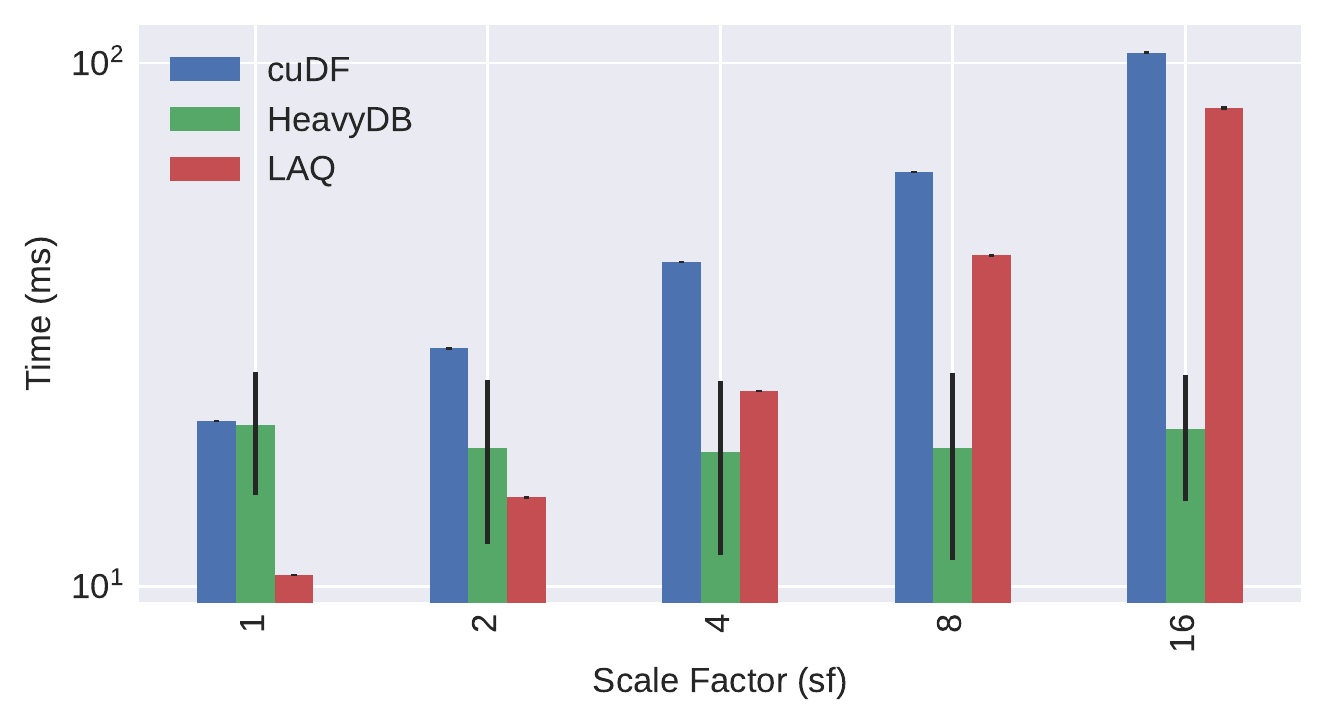}
\caption{Average execution time under various scale factors.}
\label{fig:q1}
\end{figure}

\para{Observation}
Figure \ref{fig:q1} illustrates the average execution time of all queries under different scale factors. HeavyDB presents similar average performance with different $sf$ but has significant standard errors. MM-Join exhibits a significant advantage against the other two systems at small scale factor.s As the scale factor increases, the performance of LAQ turns out to be slower than HeavyDB and approaches cuDF.

\para{Analysis}
HeavyDB is a well-designed data management system with dedicated caching mechanisms. When the evaluation executes repeatedly, more data are cached in global memory, which leads to superior performance at large $sf$. cuDF and LAQ are vanilla implementation join algorithms. They can not obtain advantages through caching strategies during repeated experiments. Another notable finding is that performance of LAQ degrades faster than cuDF due to the high computational complexity of the spMM kernel.

\mybox{$Q_2:$ How do speedups of LAQ against cuDF and HeavyDB vary w.r.t different queries?}
\begin{figure}[t!]
\centering
\includegraphics[width=2.8in]{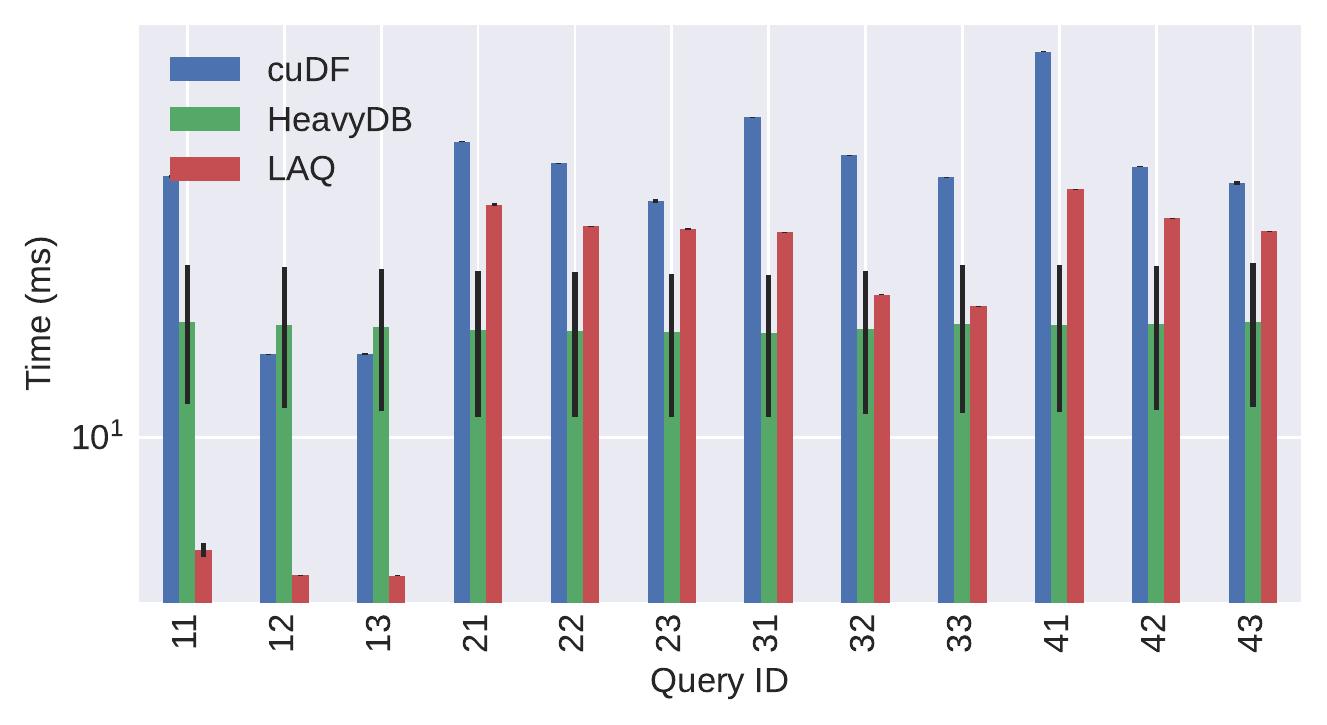}
\caption{Average execution time of different queries when $sf=4$.}
\label{fig:q21}
\end{figure}
\begin{figure}[t!]
\centering
\includegraphics[width=2.8in]{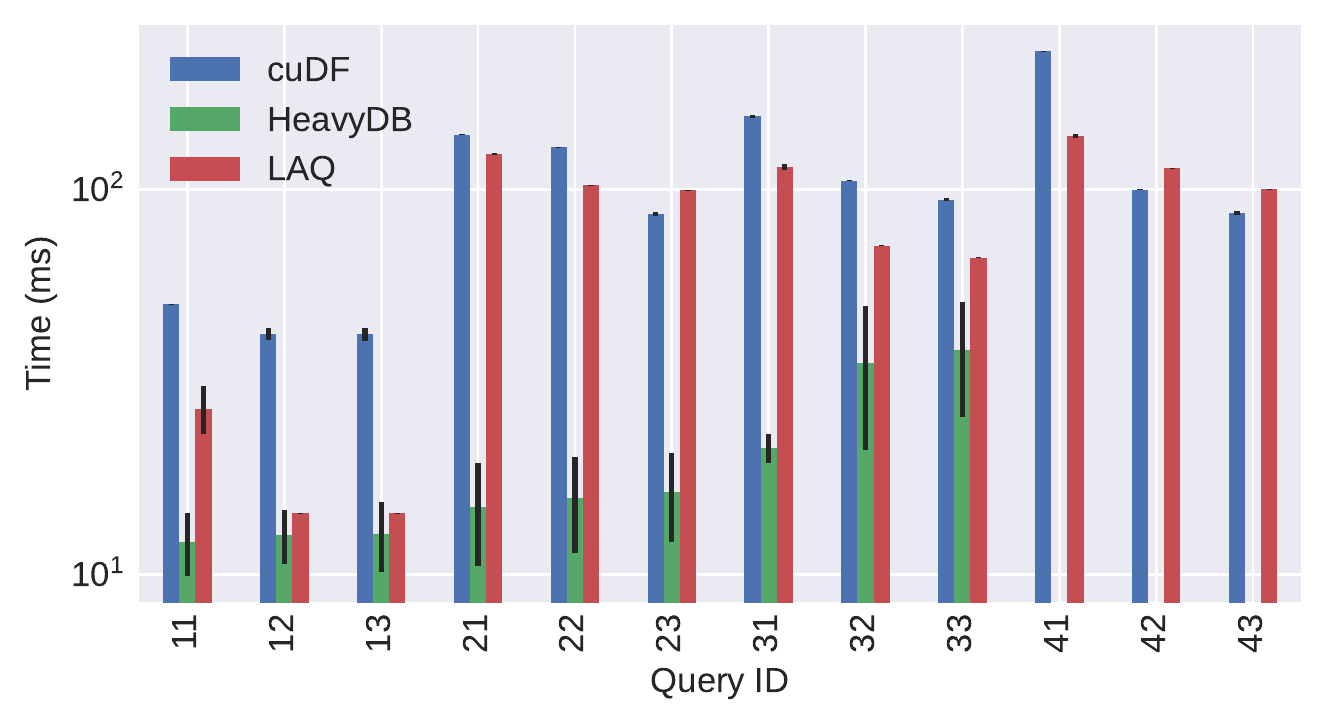}
\caption{Average execution time of different queries when $sf=16$.}
\label{fig:q22}
\end{figure}

\para{Observation}
Figure \ref{fig:q21} and \ref{fig:q22} show execution time with respect to different queries under $sf=4$ and $sf=16$. We can find out that all systems perform faster in query group 1. LAQ exhibits noticeable speedups against the other two systems in query group 1 when $sf=4$. We can also observe that LAQ becomes slower than cuDF in query 42 and 43 when $sf=16$. 

Figure \ref{fig:q22} does not show the results of HeavyDB for query group 4 when sf=16 due to out-of-memory error raised during evaluation.

\para{Analysis}
Matrix multiplication (MM), which is the most crucial operation in LAQ, is known to effectively exploit GPU parallelism due to the inherent nature of LA algorithms. However, this does not eliminate the computational complexity disadvantage of MM. When processing large-scale data, this increased complexity causes MM-Join to underperform compared to the partitioned hash join in cuDF.

Moreover, we observed a positive correlation between the performance of LAQ and the selectivity of the queries, as illustrated in Figure \ref{fig:sels}. Among all results, the performance on query 33 is exceptional. Although query 33 has lower selectivity, both algorithms exhibit slower performance due to an additional join operation. Interestingly, HeavyDB does not display a correlation between performance and selectivity because of its cache management. In query group 3, HeavyDB shows performance degradation, which can be explained by the overhead of cache eviction due to increased intermediate results generated by joins.

\mybox{$Q_3:$ Which operator in a query needs more optimization?}
\begin{figure}[h!]
\centering
\includegraphics[width=2.5in]{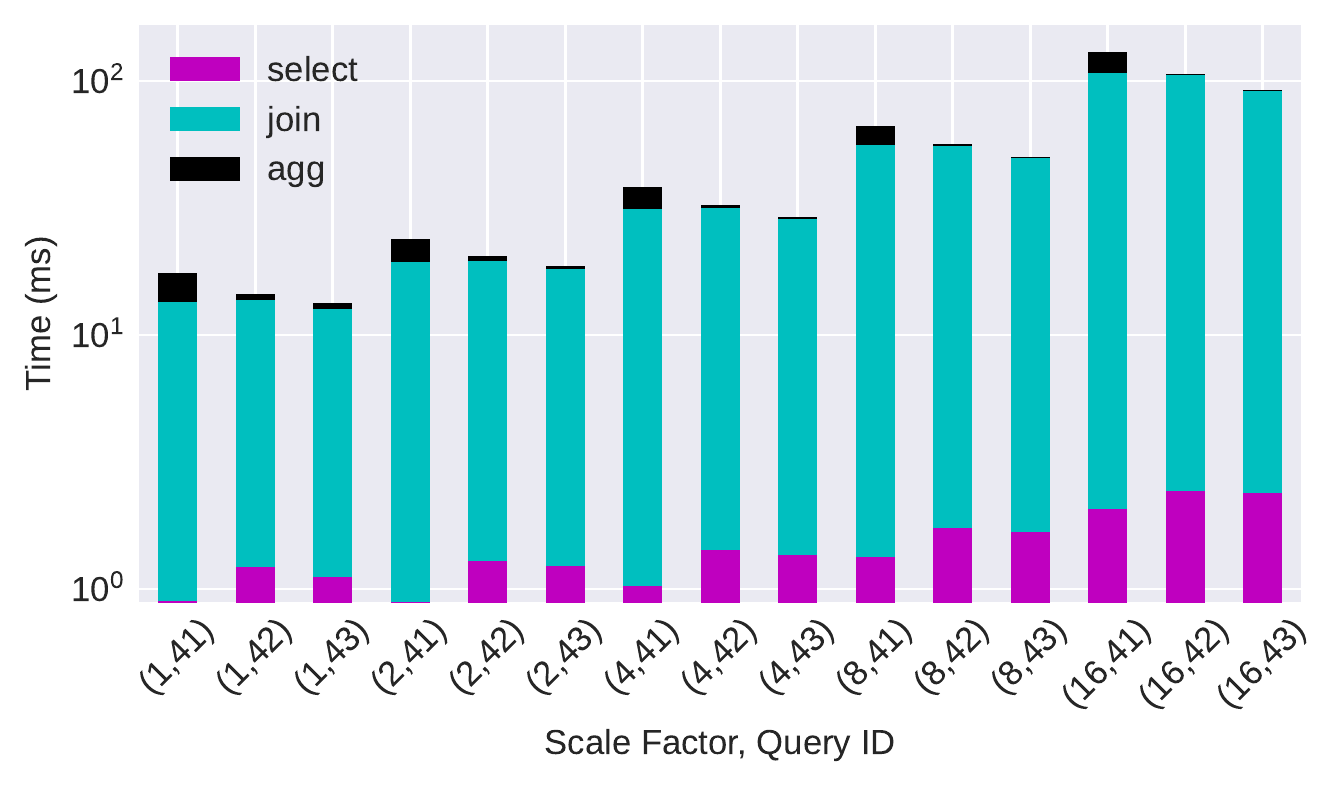}
\caption{Performance breakdown for queries w.r.t different scale factors. We take query group 4 as an example.}
\label{fig:q31}
\end{figure}
\begin{figure}[h!]
\centering
\includegraphics[width=2.5in]{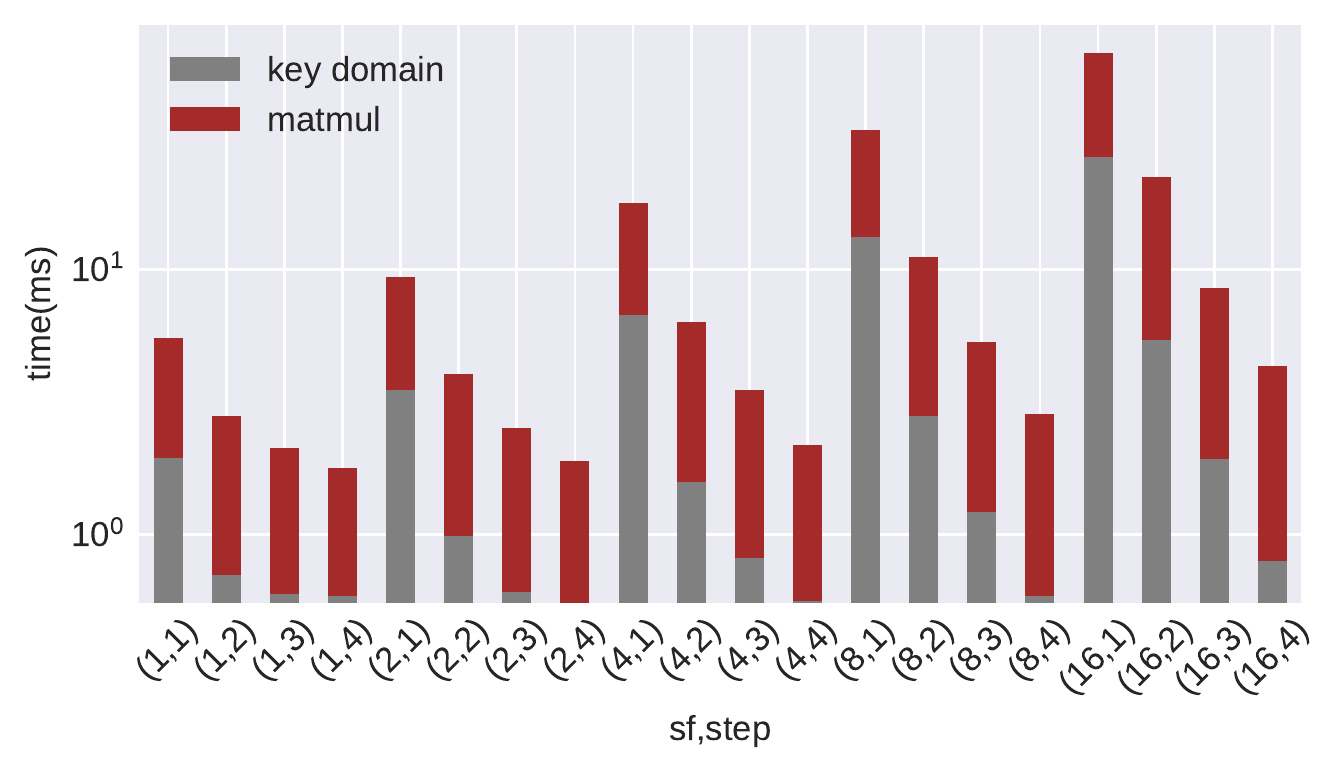}
\caption{Performance breakdown for MM-Join w.r.t different scale factors. We take query group 4 as an example. }
\label{fig:q32}
\end{figure}

\para{Observation}
In this experiment, we use query group 4 as an example to present the breakdown performance of MM-Join. Figure \ref{fig:q31} shows the execution time of different operations in a query. Apparently, join operations dominates the execution time. In Figure \ref{fig:q32}, we further investigate two primary operations of joins. Domain generations take a similar portion of execution time within 4 join steps in query group 4, whereas join operations' portion decreases as the selectivity decreases. 

\para{Analysis} In Section \ref{equijoin}, we have learned that the computational complexity of domain generation is $O(n^2\log n)$, independent of selectivities. This operation becomes particularly costly when selectivity is low. Nonetheless, the domain consists of a union of tables, allowing us to cache the domain for reuse. Caching proves advantageous when key updates are infrequent. If updates to the cached domain are necessary, the complexity of searching and inserting into a sorted array is $O(n + \log n)$, which is still more efficient than rebuilding the domain from scratch. As a result, we can further enhance performance by employing domain caching strategies.


\vspace{-1mm}
\subsection{Performance Evaluation for Operator Fusion}
\label{exp_fusing}
In this section, we evaluate operator fusion with linear operators in Section \ref{case_linear} and \ref{case_tree} to demonstrate the performance improvement that operator fusion brings to predictive pipelines. In addition to evaluating performance with different $sf$, we also examine the impact of model shape. Specifically, we vary the shape of models with different values of $k$ and $l$ to validate a potential factor, $\frac{k}{l}$, that may influence the speedup of operator fusion.

\vspace{-0.1cm}
\subsubsection{Simple Linear Operator}
This example exhibits a scenario where the output of join operations is fed to a linear operator producing a matrix. We separately evaluate two conditions: input with large $sf$, where the cardinality setting 1 is enabled, followed by a small linear operator, and input with small $sf$ (cardinality setting 2) connected to a relatively large operator. 


\mybox{$Q_{4}:$ How much speedup can operator fusion deliver in scenarios with cardinality setting 1 followed by a simple linear operator?}

\begin{figure}[h!]
\centering
\includegraphics[width=2.3in]{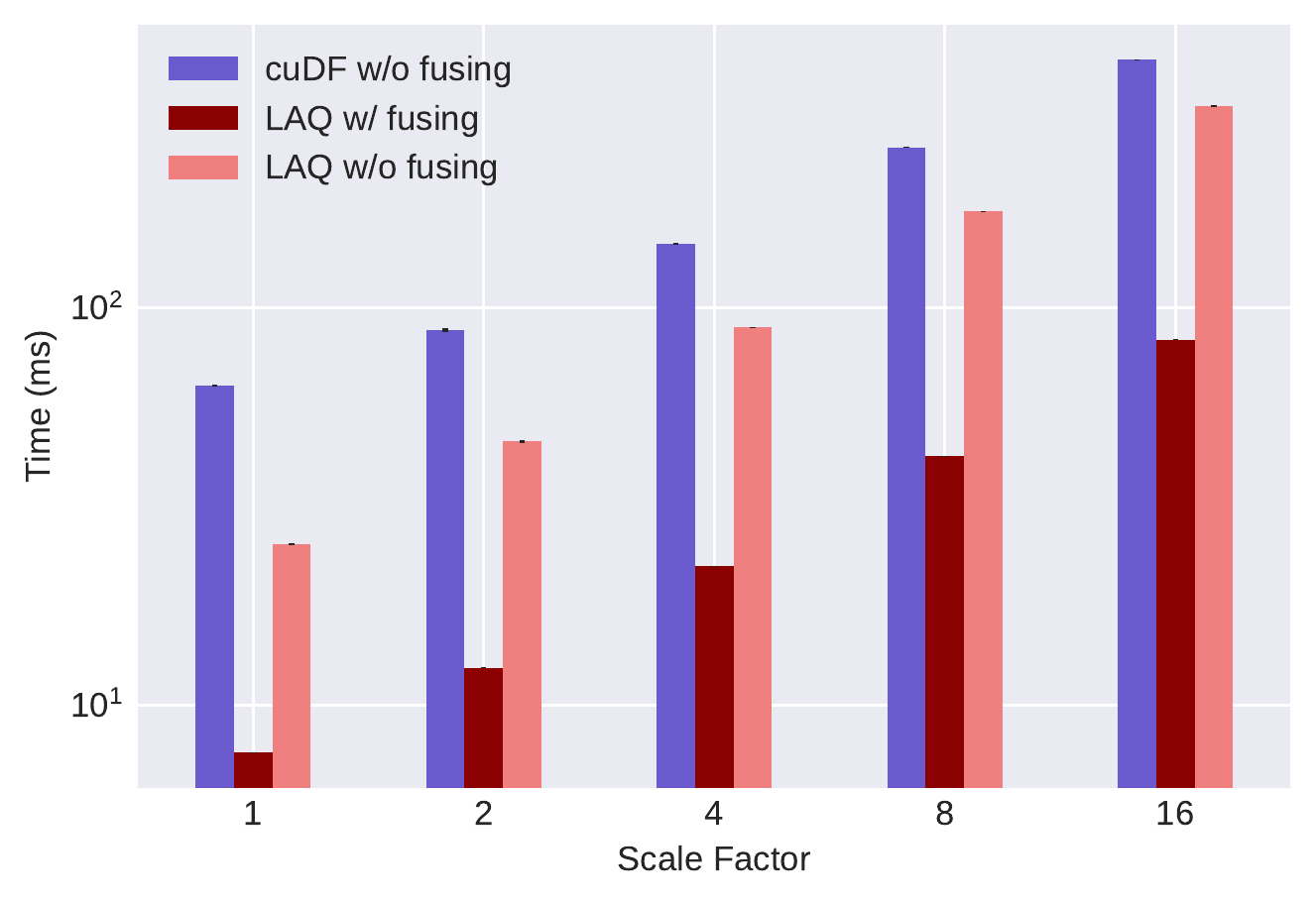}
\caption{Average execution time of join with and w/o fusing linear operators under different scale factors.}
\label{fig:q611}
\end{figure}
\begin{figure}[h!]
\centering
\includegraphics[width=2.3in]{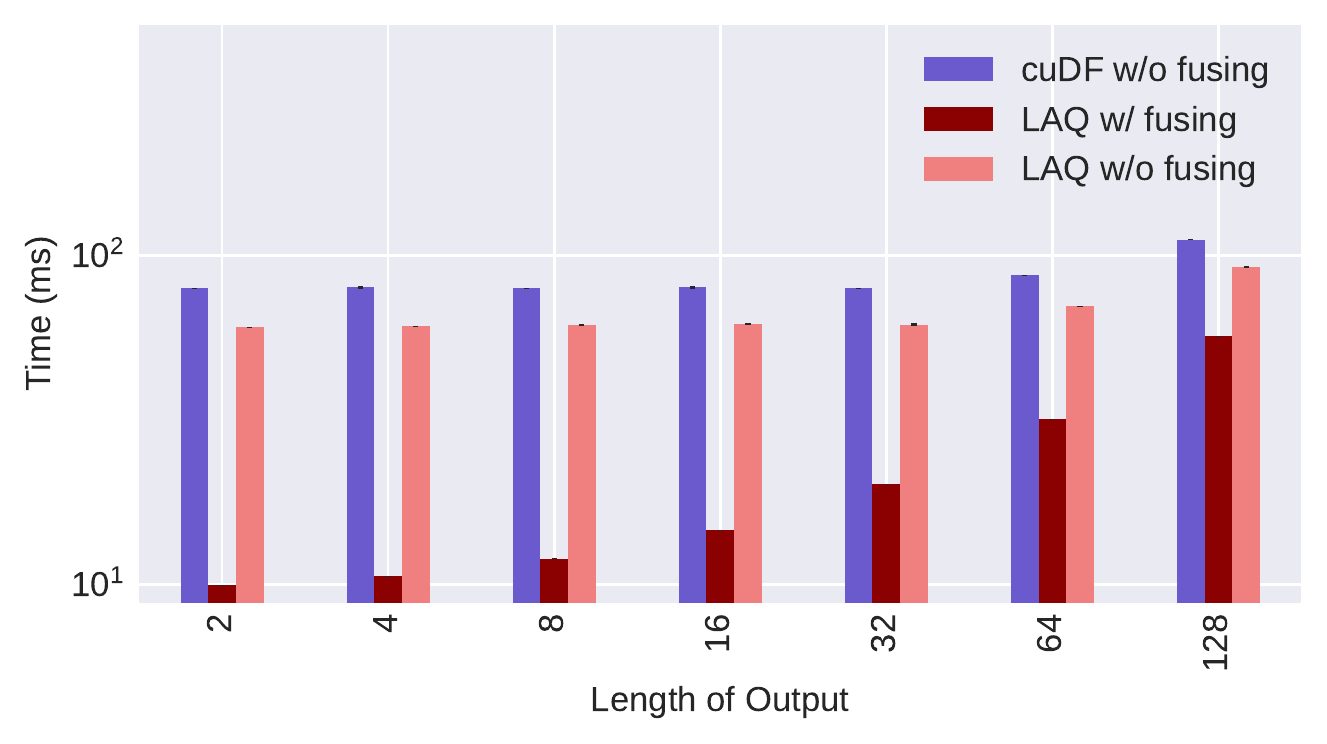}
\caption{Average execution time of predictive pipeline of simple linear operator with and w/o operator fusion when $sf=4$. The experimental scenario is large input with small model.}
\label{fig:q612}
\end{figure}
\vspace{-0.1cm}
\begin{figure}[h!]
\centering
\includegraphics[width=2.3in]{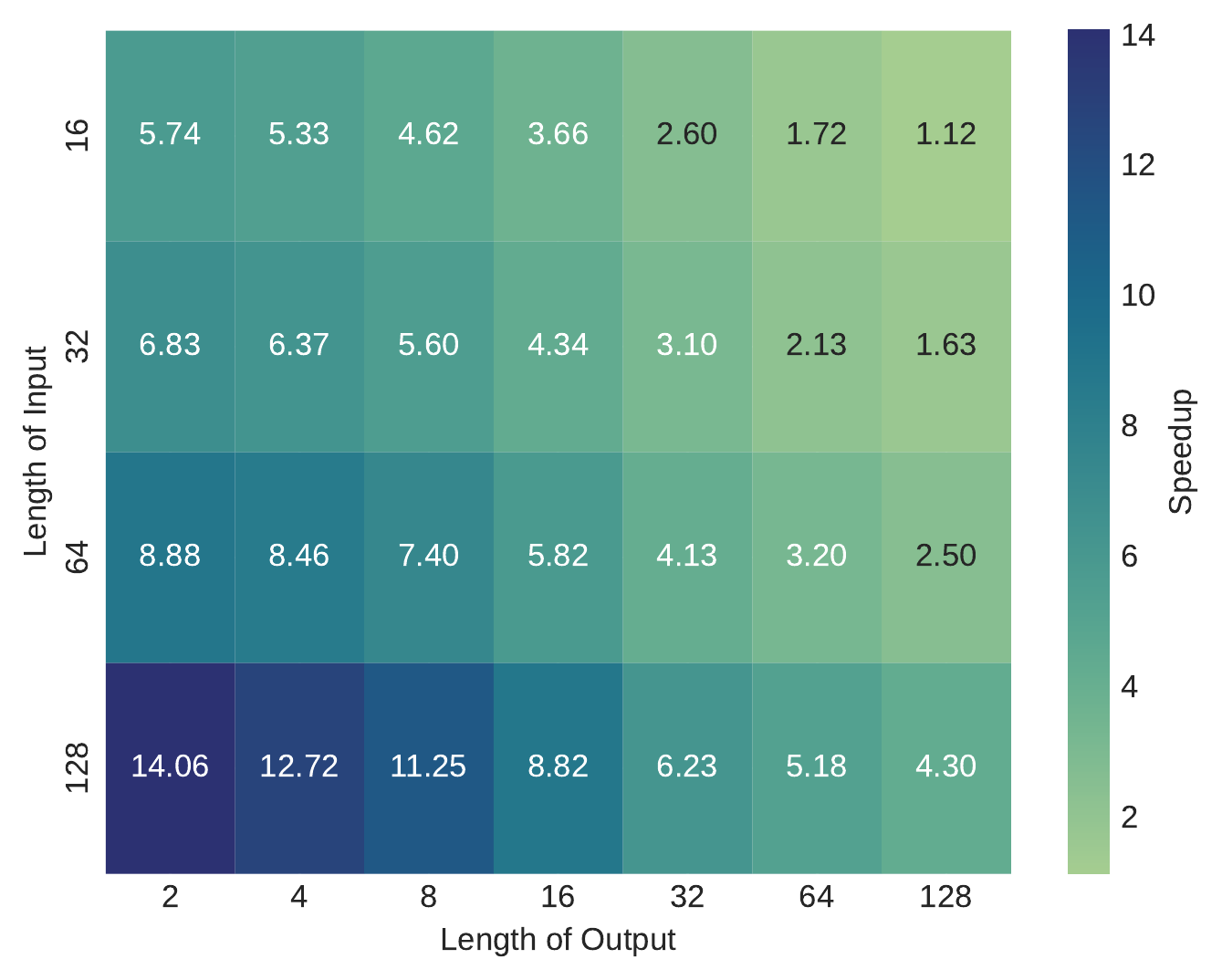}
\caption{Heatmap of speedup w.r.t lengths of input and output when $sf=8$.}
\label{fig:q613}
\end{figure}

\para{Observation}
In this experiment, we compare the execution time of a star join with operator fusion to LA-after-join implementations. It is important to note that HeavyDB is not included here, as it is slower by around 5 times of cuDF. Figure \ref{fig:q611} displays the average execution time under various scale factors. The fusion method outperforms the other two implementations. In Figure \ref{fig:q612}, we hold all parameters constant except for the output shape of the linear operator. Both cuDF and the non-fusion method do not exhibit significant changes in execution time compared to the fusion method. Although the fusion method still demonstrates speedups, these speedups continue to decrease as the output shape grows larger.

\para{Analysis} Through Equation \ref{linear_complexity}, we understand that the speedup is negatively correlated with output shape $l$ and positively correlated with input width $k$. Due to the large input size in this experiment, the lower order terms $\frac{k^2}{3il}$ and $\frac{k}{\sum_j{r_j}}$ in the equation are neglectable. Consequently, in Figure \ref{fig:q612}, we observe that the speedup of the fusion method gradually decreases as $l$ increases. Additionally, we illustrate the speedup values concerning different $k$ and $l$, while maintaining $sf=8$, in Figure \ref{fig:q613}. The highest speedup occurs at the largest $k$ and smallest $l$, whereas the lowest speedup is found along the diagonal. This result validates our analysis derived from Equation \ref{linear_complexity}.


\begin{figure}[t!]
\centering
\includegraphics[width=2.3in]{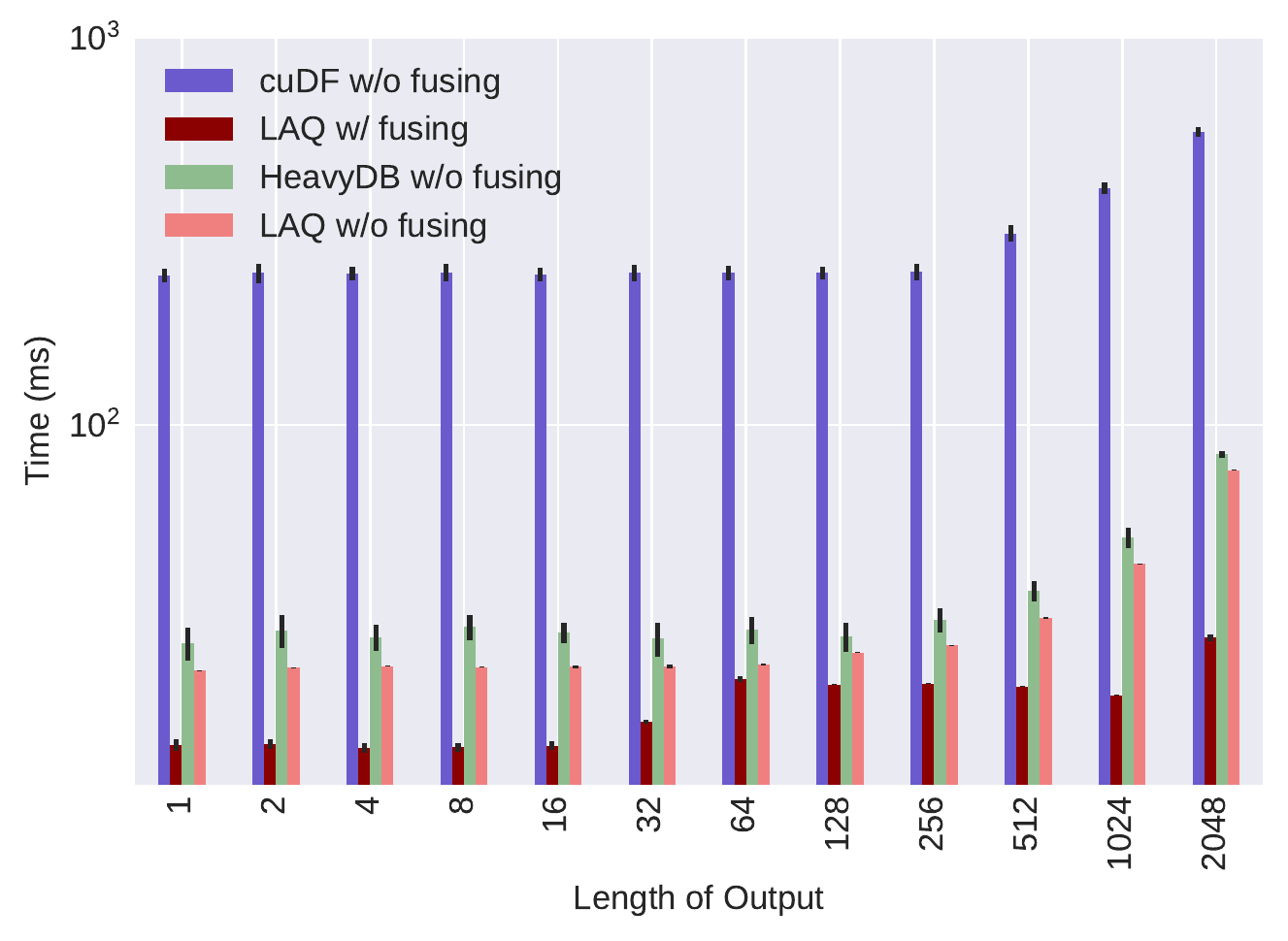}
\caption{Average execution time of predictive pipeline of simple linear operator with and w/o operator fusion when $sf=4$. The experimental scenario is small input with large model.}
\label{fig:q621}
\end{figure}
\mybox{$Q_{5}:$ How much speedup can operator fusion deliver in scenarios with cardinality setting 2 followed by a large linear operator?}

\para{Observation}
In this experiment, we set the base cardinality to 1/1000 of SSB and enlarge the output shape $l$ up to $2^{11}$. Due to the size reduction of source tables, HeavyDB also exhibits comparable performance against others. Comparing Figure~\ref{fig:q621} and \ref{fig:q612}, we can clearly observe much more significant speedups of operator fusion in small dimension tables in cardinality setting 2.

\para{Analysis}
As indicated by Equation \ref{linear_complexity}, a reduction in input cardinality corresponds to a smaller value for $\sum_j{r_j}$, resulting in a larger speedup. Furthermore, HeavyDB is slower than both LAQ with and without operator fusion due to data structure conversions across different runtimes between the database and ML systems. Thus, we can conclude that the fusion method is more advantageous when processing linear queries with small dimension tables.

\mybox{$Q_{6}:$ How much time does the pre-fusion phase take?}
\begin{figure}[h!]
\centering
\includegraphics[width=2.8in]{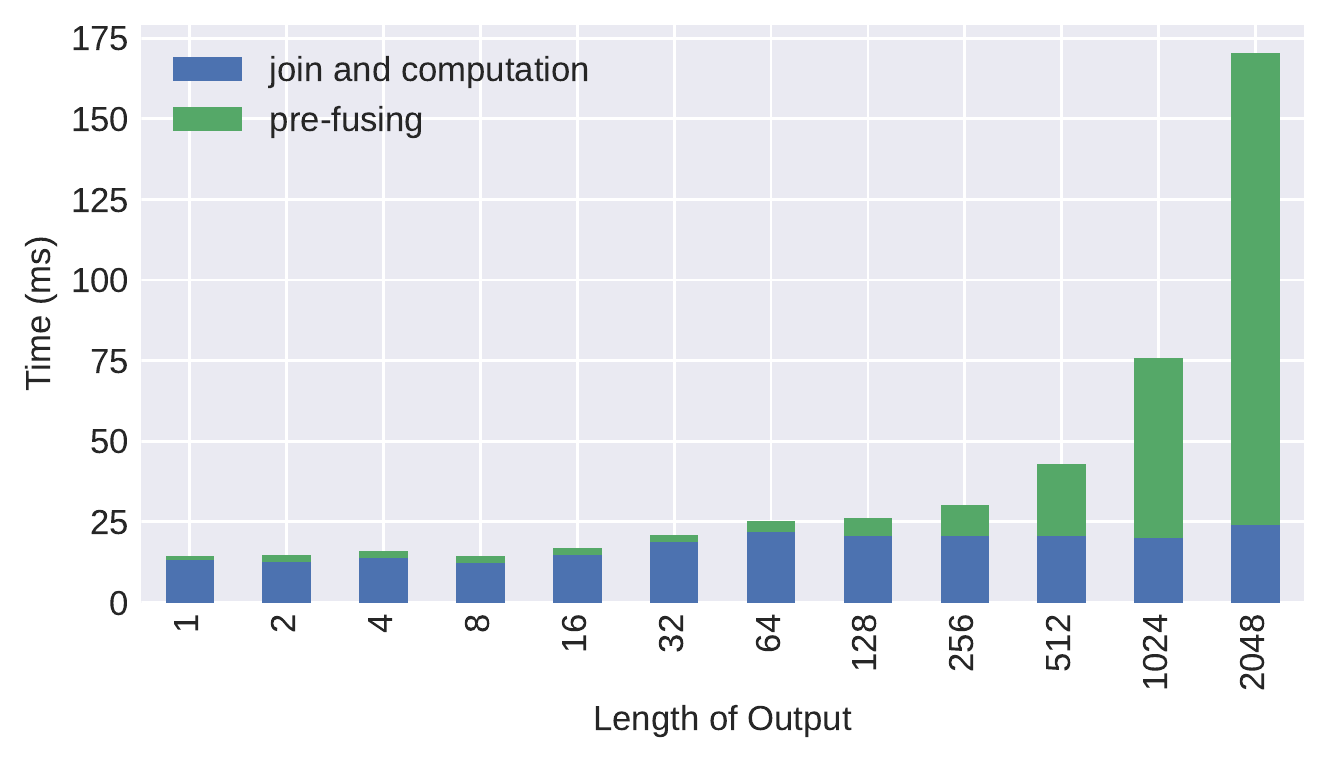}
\caption{The execution time that pre-fusion stage and join-computation stage take in prediction with linear operator after joining.  }
\label{fig:q631}
\end{figure}

\para{Observation \& Analysis}
While the operator fusion method provides considerable speedup, it is crucial to consider the cost of the pre-fusion step, as shown in the underlined parts of \ref{eq:linear_op_origin}. This is because dimension tables, although updated less frequently than fact tables, are not static constants. Moreover, the pre-fused tables may be larger than the original dimension tables when the output shape exceeds the number of columns, resulting in increased memory usage. Consequently, a quantitative trade-off between fusion and non-fusion methods still calls for further study in practice.

Figure \ref{fig:q631} presents a stacked plot illustrating the relative proportion between pre-fusion and subsequent multiplication with $I_*$. Based on the parameter settings in Q5, we observe that when the output shape $l$ is less than or equal to 512, the linear operation dominates the total execution time. As a result, if memory constraints are present, we can prioritize query completion without encountering out-of-memory errors, considering the diminishing speedup with larger output shapes.
\vspace{-0.1cm}
\subsubsection{Decision Tree}
In this experiment, we substitute the simple linear operator with a more intricate decision tree model to explore the performance advantages resulting from operator fusion. Following a similar experimental approach for simple linear operators, we separately assess the performance of two scenarios: cardinality setting 1 followed by a simple decision tree and cardinality setting 2 followed by a relatively large model.

\mybox{$Q_{7}:$  How much speedup can operator fusion deliver in scenarios with cardinality setting 1 followed by a simple decision tree?}
\begin{figure}[t!]
\centering
\includegraphics[width=2.3in]{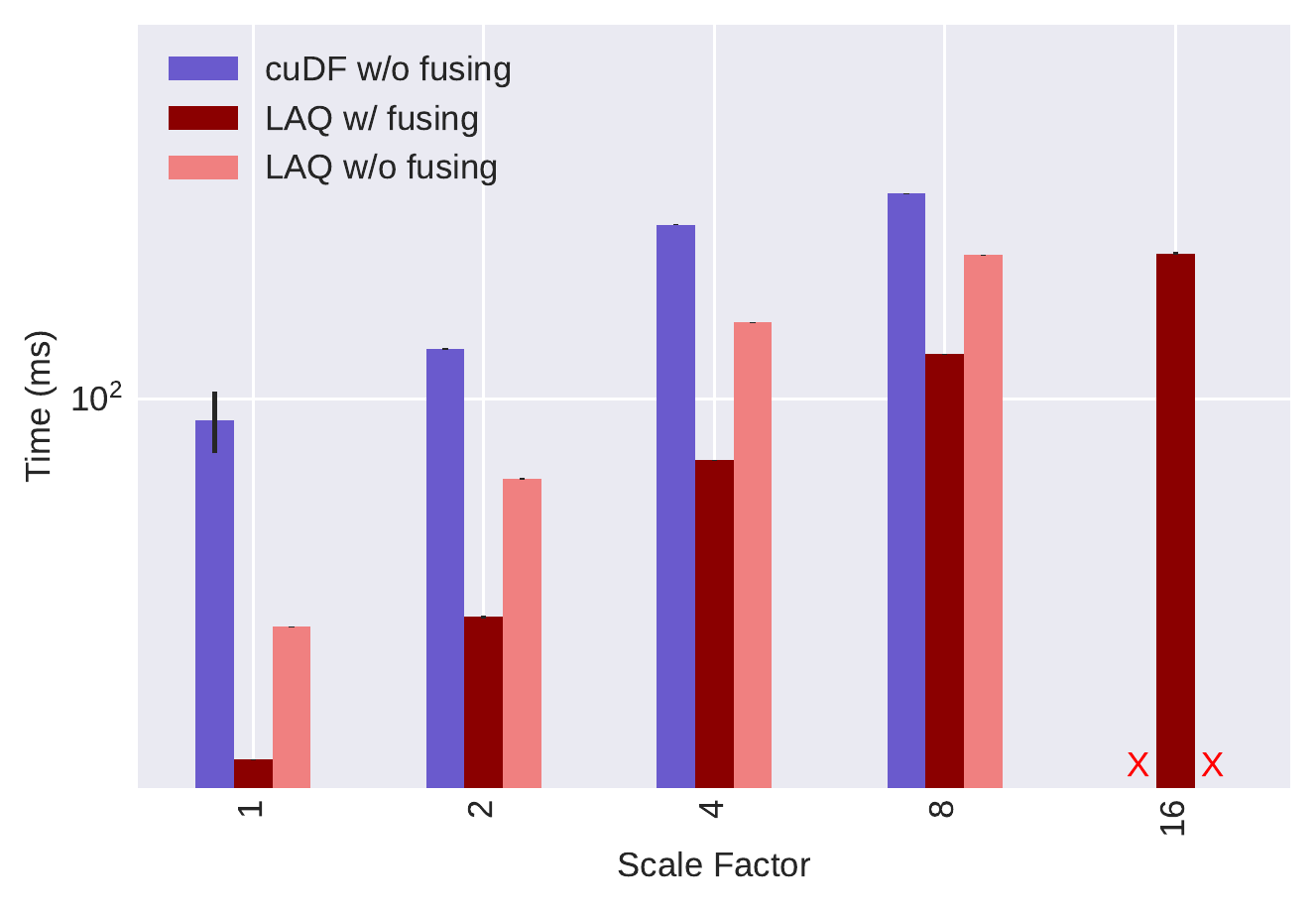}
\caption{Average execution time of join with and w/o fusing decision trees under different scale factors.}
\label{fig:q711}
\end{figure}
\begin{figure}[t!]
\centering
\includegraphics[width=2.3in]{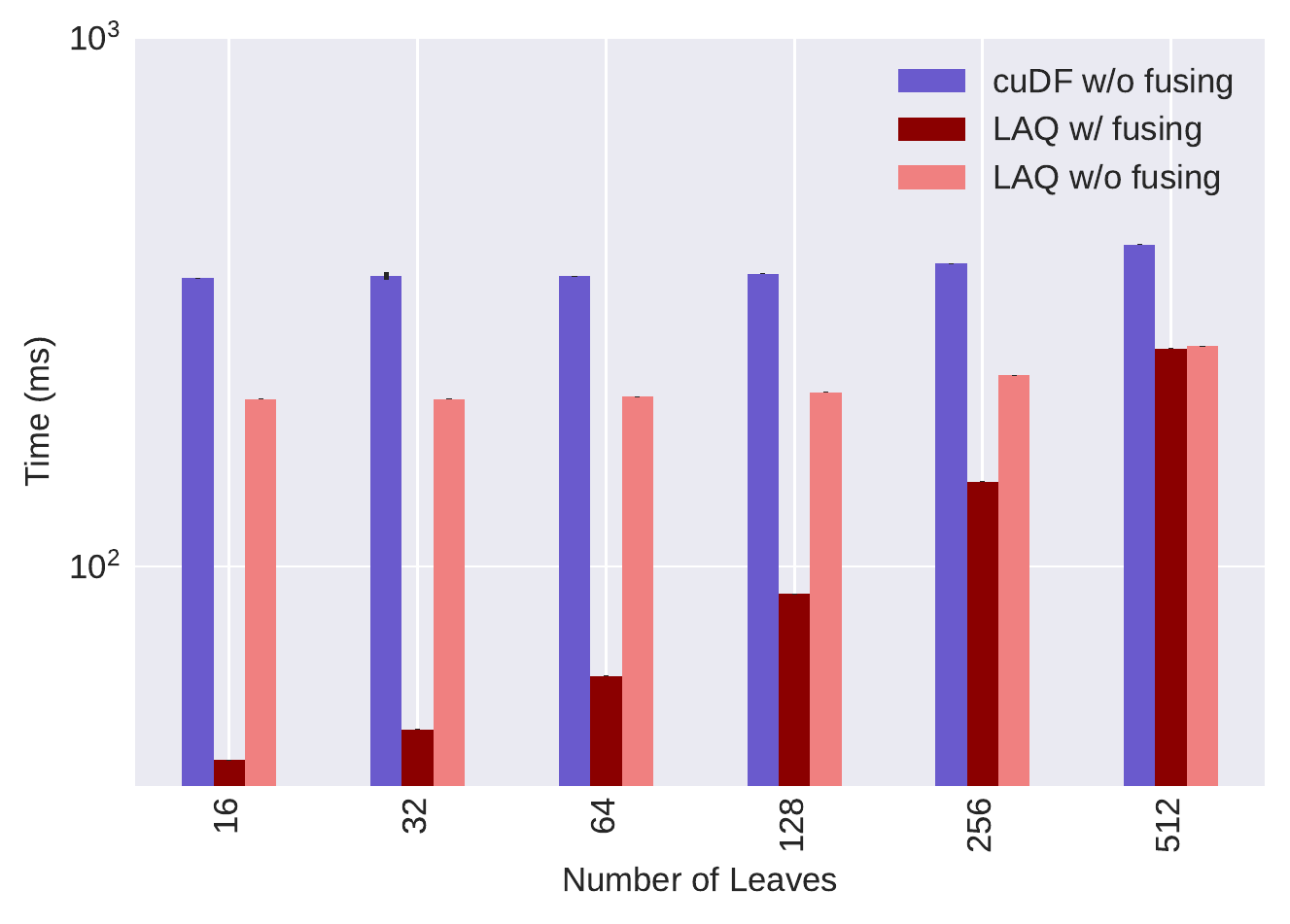}
\caption{Average execution time of predictive pipeline of decision tree with and w/o operator fusion when $sf=4$. The experimental scenario is large input with small model.}
\label{fig:q712}
\end{figure}
\begin{figure}[t!]
\centering
\includegraphics[width=2.3in]{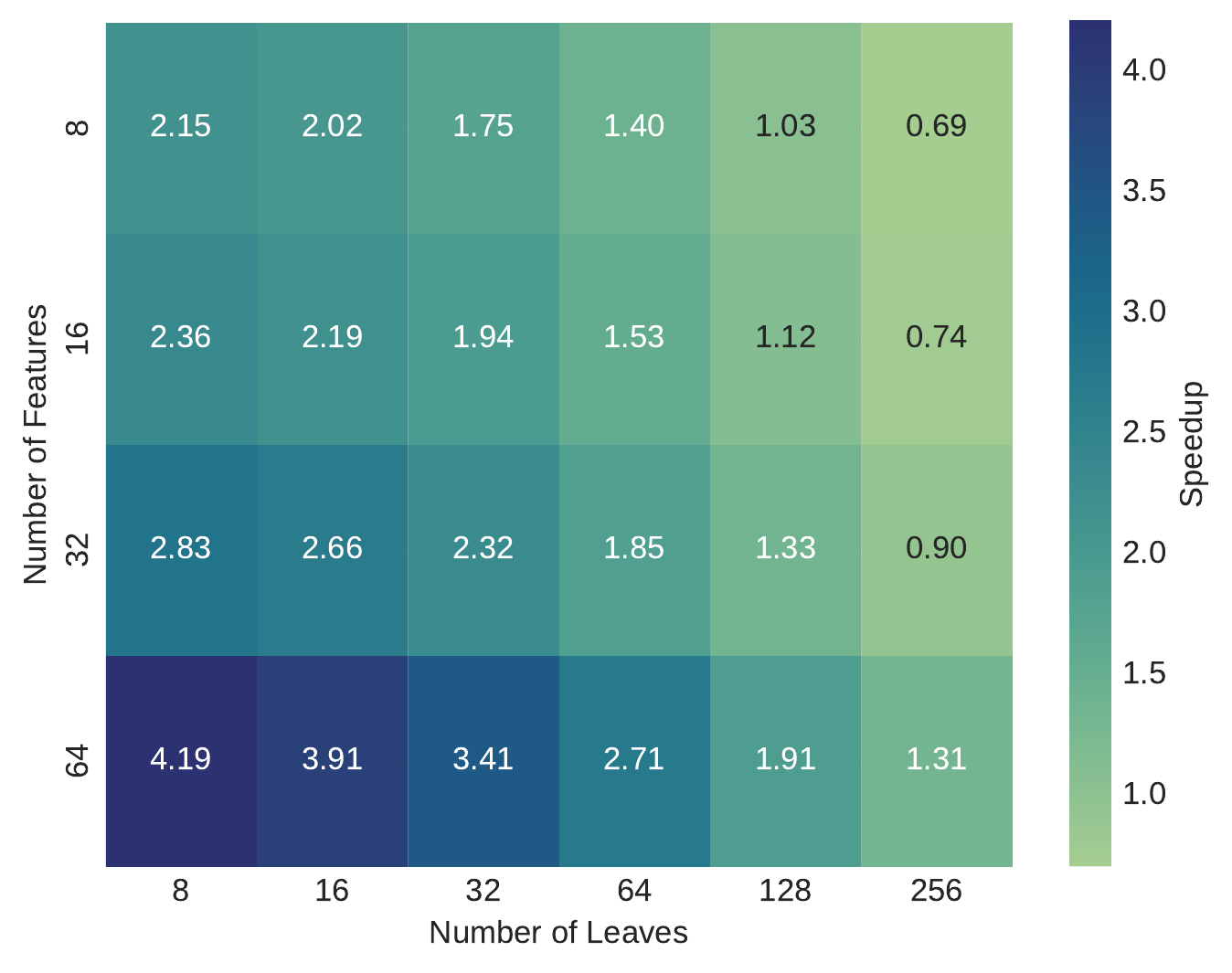}
\caption{Heatmap of speedup w.r.t numbers of features and leaves when $sf=8$.}
\label{fig:q713}
\end{figure}
\vspace{0.1cm}

\para{Observation \& Analysis}
Figure \ref{fig:q711}-\ref{fig:q713} display the results for large input scenarios. Figure \ref{fig:q711} demonstrates that the average execution time of the fusion method is significantly faster than the other two methods across all scale factors. Notably, both cuDF and the non-fusion method fail to execute due to out-of-memory errors, while the fusion method completes a larger portion of evaluations. In Figure \ref{fig:q712}, we vary parameter $l$ while keeping $sf=4$. It is evident that LAQ with fusion outperforms other methods when $l$ is low, but its performance deteriorates as $l$ increases.

The performance degradation can be explained using Equation \ref{tree_comp_frac}. We focus on $\frac{k}{l}$ because the remaining terms can be disregarded with large $\sum_j{r_j}$. As $l$ increases, $\frac{k}{l}$ decreases, leading to a reduced speedup compared to the non-fusion method. In Figure \ref{fig:q713}, we examine the speedup concerning different values of $k$ and $l$. The highest speedup occurs at the largest $k$ and smallest $l$, which validates our complexity analysis that the speedup is correlated with $\frac{k}{l}$. A large $k$ and small $l$ suggest that the model functions as a data compressor, indicating that the fusion method can be advantageous when applying a narrow-down model to a large amount of data. From a hardware perspective, a pre-fusion method with a filtering effect actually reduces the size of input data, which further decreases memory usage and memory I/O in subsequent computations. Therefore, the value of $\frac{k}{l}$ can serve as a potential indicator for determining whether pre-fusion should be applied.

\mybox{$Q_{8}:$ How much speedup can operator fusion deliver in scenarios with cardinality setting 2 followed by a large decision tree?}
\begin{figure}[t!]
\centering
\includegraphics[width=2.8in]{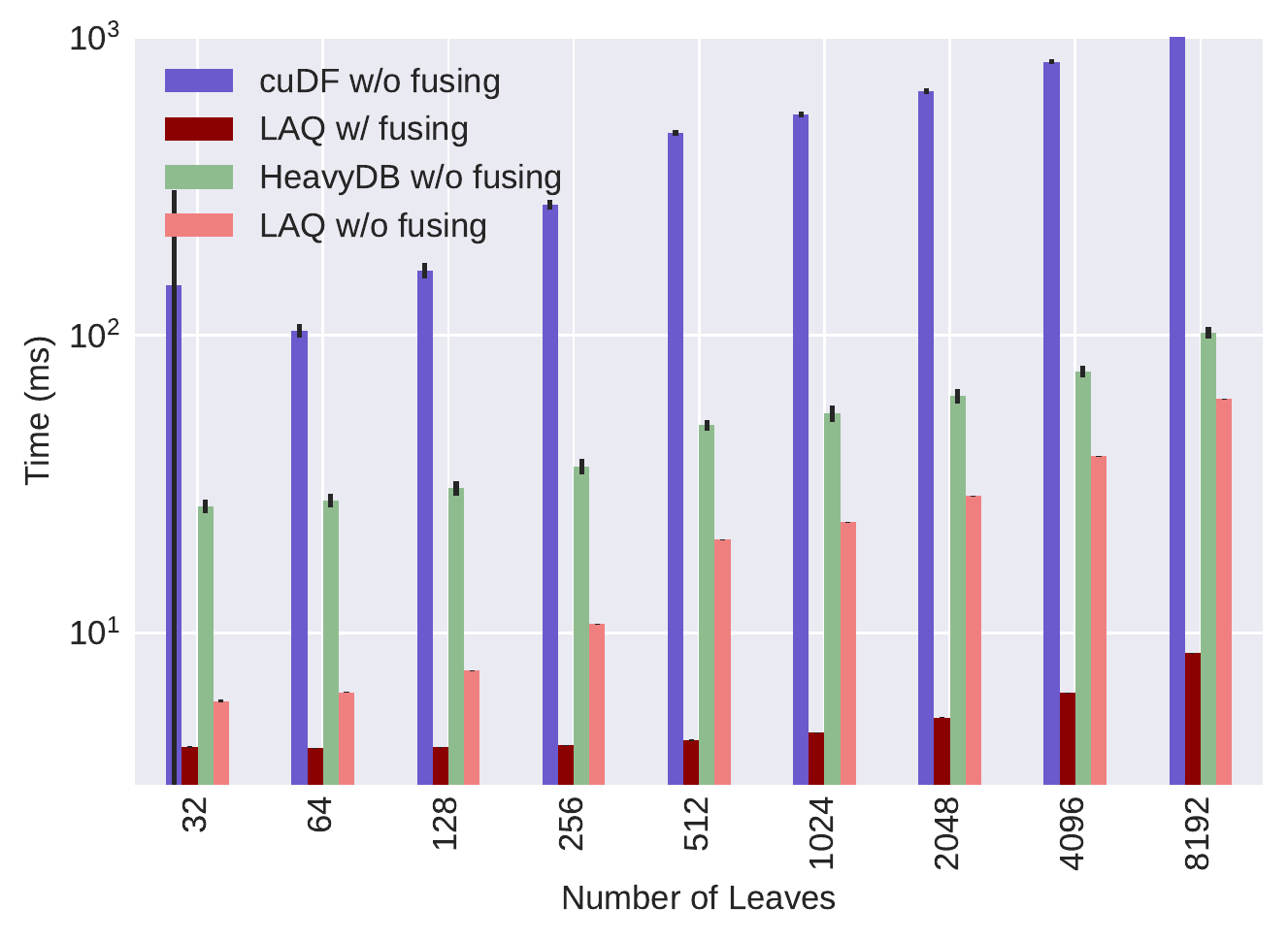}
\caption{Average execution time of predictive pipeline of decision tree with and w/o operator fusion when $sf=4$. The experimental scenario is small input with large model.}
\label{fig:q721}
\end{figure}
\begin{figure}[h!]
\centering
\includegraphics[width=2.8in]{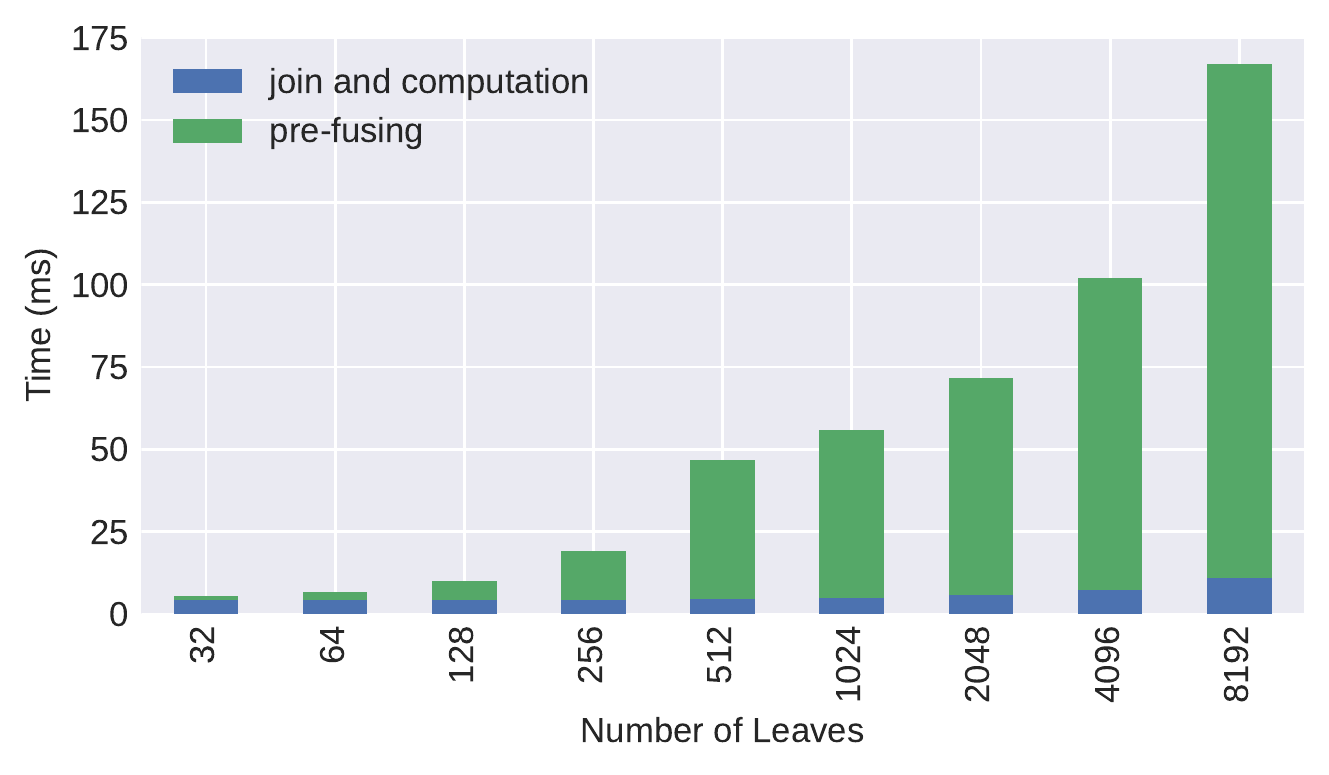}
\caption{The execution time that pre-fusing stage and join-computation stage take in prediction with decision tree after joining.}
\label{fig:q731}
\end{figure}

\para{Observation \& Analysis}
In scenarios where dimension tables with small cardinality are processed using a large model, the operator fusion method exhibits a more significant speedup compared to the other three methods, as illustrated in Figure~\ref{fig:q721}. When the input scale factor is reduced to 1\% of that in Q8, the residual terms  in Equation~\ref{tree_comp_frac} can no longer be ignored, leading to a greater speedup. However, as the model size increases, the cost of pre-operator fusion becomes more expensive relative to subsequent computations, as shown in Figure \ref{fig:q731}. Considering that dimension tables are not entirely static but updated according to changes in the dimension tables, the actual benefits of the operator fusion method depend on the update frequency of the dimension tables.

\section{Related work}
\label{sec:rw}

\para{GPU relational data processing}
GPU-accelerated query processing has been extensively researched in recent decades. As GPU architectural design and memory bandwidth between hosts and GPUs have advanced, several database management systems (DBMS) have incorporated GPU acceleration to optimize their query processing capabilities. Notable examples of GPU-based systems include Crystal, OmniSci (now known as HeavyDB)~\cite{heavy}, BlazingSQL \cite{blazing}, and PG-Strom. These systems take advantage of the parallel processing capabilities of GPUs to perform operations such as filtering, aggregation, and join processing at a significantly faster rate compared to traditional CPU-based systems. 

However, these works do not change the nature of relational data processing. The theoretical and practical gap between relational data and linear algebraic input for machine learning still hinders potential integration and optimization opportunities.


\para{Query processing using matrix multiplication}
Matrix multiplication has been widely adopted in graph query processing. Earlier research \cite{joinproject1,joinproject2} proposed LA-based algorithms for computing an equi-join followed by a duplicate-eliminating projection, which yields smaller intermediate results and more efficient memory I/O than conventional relational operators. One recent paper \cite{joinproject3} proposed \emph{DIM3} to address several performance bottlenecks in \cite{joinproject2}. \emph{DIM3} introduces partial result caching and support for join-aggregation operations. However, this line of research focuses primarily on join operations rather than a general method of processing relational queries with linear algebra (LAQ) discussed in our research.

TCUDB \cite{TCU} is the first GPU query engine that primarily uses LAQ as its query engine, which implements equi-join and single-column aggregation using LAQ. The design principle of the join-aggregation operator in TCUDB is similar to that of \cite{joinproject1} and \cite{joinproject2}, but it is embedded within a query planner that supports a wide range of SQL queries and analytic queries. In the TCUDB paper, the authors evaluate its performance with graph query workloads, but they do not provide insights into its performance into the cost of each operator in LAQ. In contrast, our work extensively evaluates LAQ on a wide range of data and reports detailed performance breakdown.

To support an integrated pipeline of data integration and ML model training, in our previous work \cite{hai2023amalur}, we have defined matrix-based representations for mapping columns and rows between source and target tables. With the logical representations, we identify the method to evaluate outer-join, inner-join, left-join, and union in data integration tasks using linear algebra operators. Building upon this foundation, in the current research, we evaluate the extended LAQ using relational query benchmark datasets to assess its performance in traditional data queries and predictive pipelines.

\para{Cross-optimization of ML and relational data processing}
Raven\cite{raven} and LaraDB \cite{lara} implemented cross-optimization methods for batch prediction tasks that follow relational data processing. The optimizer, built on a unified intermediate representation, enables the exchange of information between relational operators and ML models. However, in this research, the relational and linear components must execute in separate runtimes, which may involve potential data transformation and communication overhead. In contrast, our method unifies the data processing and ML model prediction in representation as well as runtime.


Hummingbird \cite{hummingbird1,hummingbird2} is a system capable of compiling a wide range of traditional ML models into modern tensor-based runtimes designed specifically for deep learning models. In addition to providing a unified runtime, Hummingbird employs deep learning compilers to optimize the overall efficiency of the ML pipeline.

However, despite the benefits of tensor representation, the operators in Hummingbird do not implement joins and aggregations commonly found in data integration and training data generation processes. 

Inspired by Hummingbird, TQP \cite{tqp} further extends tensor programs for relational operations, including sort-merge join and hash join, enabling it to handle the full TPC-H benchmark \cite{tpch}. TQP leverages a widely-used tensor computing runtime, to optimize and execute workflows containing both relational data processing and model prediction on GPUs. Following this research, TDP \cite{tdp} expands capabilities to encode multi-modal data processing. Nevertheless, the physical implementation of join and aggregation operators remains in the relational style rather than LA. This diversity prevents the differentiability from being further pushed down to the source data before joins and also misses optimization opportunities brought about by LA rewriting. Our research implements joins and aggregations in linear algebra and proposes an operator fusion method leveraging this unified theoretical language, significantly accelerating predictive pipelines.


\vspace{-2mm}
\section{Conclusion and Future Research}
In this paper, we present the operator fusion method to optimize the speed of predictive pipelines consisting of data processing and ML model predictions. By employing LAQ to represent data query processing, our approach can merge operators in ML model predictions with data processing operators. Furthermore, through the analysis of the complexity of operator fusion and LAQ without operator fusion, we find that the length ratio of input vector and output vector, described as $\frac{k}{l}$ as discussed in Section \ref{fusing_method}, may influence the speedup of our method in the context of the star schema. In our evaluation, we use a widely-adopted data query benchmark, SSB, and a synthetic dataset to test the performance of LAQ and operator fusion. Based on the experimental results, we draw the following conclusions:
\begin{itemize}
    \item LAQ outperforms cuDF, a standard GPU relational query processor, in most evaluations except for query group 4 when $sf$ is 16. The inherent high computational complexity of domain construction and matrix multiplication dominates the execution time, causing performance degradation when data sizes increase. However, we can expect performance improvement by caching key domains.

    \item In experiments for predictive pipelines in Section \ref{exp_fusing}, operator fusion exhibits significant speedups up to 317x compared to the LAQ without operator fusion. Moreover, the experiment results confirm the hypothesis that $\frac{k}{l}$ in Equation \ref{linear_complexity} and \ref{tree_comp_frac} affect the speedup of operator fusion through.

    \item The speedup of operator fusion also depends on the sizes of input matrices. Fusing large models is costly, but it can be beneficial when the update frequencies and cardinality of dimension tables are low. We need to make trade-offs between operator fusion and non-operator fusion based on update patterns and data sizes.
\end{itemize}

\para{Future research} Based on the observations and analysis from the experiments, we identify that LAQ and optimizations in integrated data processing and ML pipelines call for further research.

Although we have preliminarily shown that fusing linear operators in ML models with LAQ is beneficial, a detailed cost estimation that can assist with automatic pipeline optimization is still missing. Furthermore, in the context of thriving large-scale deep learning, more operator fusion rules that can optimize deep learning operators are urgently needed. Last but not least, exploring the optimization of training performance with similar linear algebraic operator fusion techniques is also a valuable research direction.

\bibliographystyle{ACM-Reference-Format}
\bibliography{main}


\begin{thebibliography}{26}


\ifx \showCODEN    \undefined \def \showCODEN     #1{\unskip}     \fi
\ifx \showDOI      \undefined \def \showDOI       #1{#1}\fi
\ifx \showISBNx    \undefined \def \showISBNx     #1{\unskip}     \fi
\ifx \showISBNxiii \undefined \def \showISBNxiii  #1{\unskip}     \fi
\ifx \showISSN     \undefined \def \showISSN      #1{\unskip}     \fi
\ifx \showLCCN     \undefined \def \showLCCN      #1{\unskip}     \fi
\ifx \shownote     \undefined \def \shownote      #1{#1}          \fi
\ifx \showarticletitle \undefined \def \showarticletitle #1{#1}   \fi
\ifx \showURL      \undefined \def \showURL       {\relax}        \fi
\providecommand\bibfield[2]{#2}
\providecommand\bibinfo[2]{#2}
\providecommand\natexlab[1]{#1}
\providecommand\showeprint[2][]{arXiv:#2}

\bibitem[Amossen and Pagh(2009)]%
        {joinproject1}
\bibfield{author}{\bibinfo{person}{Rasmus~Resen Amossen} {and}
  \bibinfo{person}{Rasmus Pagh}.} \bibinfo{year}{2009}\natexlab{}.
\newblock \showarticletitle{{Faster Join-Projects and Sparse Matrix
  Multiplications}}. In \bibinfo{booktitle}{\emph{ICDT 2009}} (St. Petersburg,
  Russia). \bibinfo{publisher}{Association for Computing Machinery},
  \bibinfo{address}{New York, NY, USA}, \bibinfo{pages}{121–126}.
\newblock
\urldef\tempurl%
\url{https://doi.org/10.1145/1514894.1514909}
\showDOI{\tempurl}


\bibitem[Balkesen et~al.(2013)]%
        {parallelradix}
\bibfield{author}{\bibinfo{person}{Cagri Balkesen}, \bibinfo{person}{Jens
  Teubner}, \bibinfo{person}{Gustavo Alonso}, {and} \bibinfo{person}{M~Tamer
  {\"O}zsu}.} \bibinfo{year}{2013}\natexlab{}.
\newblock \showarticletitle{{Main-memory hash joins on multi-core CPUs: Tuning
  to the underlying hardware}}. In \bibinfo{booktitle}{\emph{ICDE 2013}}.
  \bibinfo{pages}{362--373}.
\newblock


\bibitem[{BlazingDB}(2020)]%
        {blazing}
\bibfield{author}{\bibinfo{person}{{BlazingDB}}.}
  \bibinfo{year}{{2020}}\natexlab{}.
\newblock \bibinfo{title}{{BlazingSQL}}.
\newblock
  \bibinfo{howpublished}{\url{https://github.com/BlazingDB/blazingsql}}.
\newblock


\bibitem[Chen et~al.(2018)]%
        {tvm}
\bibfield{author}{\bibinfo{person}{Tianqi Chen}, \bibinfo{person}{Thierry
  Moreau}, \bibinfo{person}{Ziheng Jiang}, \bibinfo{person}{Lianmin Zheng},
  {et~al\mbox{.}}} \bibinfo{year}{2018}\natexlab{}.
\newblock \showarticletitle{{TVM}: An Automated {End-to-End} Optimizing
  Compiler for Deep Learning}. In \bibinfo{booktitle}{\emph{OSDI 2018}}.
  \bibinfo{pages}{578--594}.
\newblock


\bibitem[Deep et~al.(2020)]%
        {joinproject2}
\bibfield{author}{\bibinfo{person}{Shaleen Deep}, \bibinfo{person}{Xiao Hu},
  {and} \bibinfo{person}{Paraschos Koutris}.} \bibinfo{year}{2020}\natexlab{}.
\newblock \showarticletitle{{Fast Join Project Query Evaluation Using Matrix
  Multiplication}}. In \bibinfo{booktitle}{\emph{SIGMOD 2020}}.
  \bibinfo{pages}{1213–1223}.
\newblock


\bibitem[Gandhi et~al.(2023)]%
        {tdp}
\bibfield{author}{\bibinfo{person}{Apurva Gandhi}, \bibinfo{person}{Yuki
  Asada}, \bibinfo{person}{Victor Fu}, \bibinfo{person}{Advitya Gemawat},
  \bibinfo{person}{Lihao Zhang}, \bibinfo{person}{Rathijit Sen},
  \bibinfo{person}{Carlo Curino}, \bibinfo{person}{Jes{\'u}s
  Camacho-Rodr{\'\i}guez}, {and} \bibinfo{person}{Matteo Interlandi}.}
  \bibinfo{year}{2023}\natexlab{}.
\newblock \showarticletitle{{The Tensor Data Platform: Towards an AI-centric
  Database System}}. In \bibinfo{booktitle}{\emph{CIDR 2023}}.
\newblock


\bibitem[Ghiran and Buchmann(2019)]%
        {datafab}
\bibfield{author}{\bibinfo{person}{Ana-Maria Ghiran} {and}
  \bibinfo{person}{Robert~Andrei Buchmann}.} \bibinfo{year}{2019}\natexlab{}.
\newblock \showarticletitle{{The Model-Driven Enterprise Data Fabric: A
  Proposal Based on Conceptual Modelling and Knowledge Graphs}}. In
  \bibinfo{booktitle}{\emph{Knowledge Science, Engineering and Management}},
  \bibfield{editor}{\bibinfo{person}{Christos Douligeris},
  \bibinfo{person}{Dimitris Karagiannis}, {and} \bibinfo{person}{Dimitris
  Apostolou}} (Eds.). \bibinfo{publisher}{Springer International Publishing},
  \bibinfo{pages}{572--583}.
\newblock


\bibitem[Hai et~al.(2023)]%
        {hai2023amalur}
\bibfield{author}{\bibinfo{person}{Rihan Hai}, \bibinfo{person}{Christos
  Koutras}, \bibinfo{person}{Andra Ionescu}, \bibinfo{person}{Ziyu Li},
  \bibinfo{person}{Wenbo Sun}, \bibinfo{person}{van~Schijndel Jessie},
  \bibinfo{person}{Yan Kang}, {and} \bibinfo{person}{Asterios Katsifodimos}.}
  \bibinfo{year}{2023}\natexlab{}.
\newblock \showarticletitle{{Amalur: Data Integration Meets Machine Learning}}.
  In \bibinfo{booktitle}{\emph{ICDE 2023}}. \bibinfo{pages}{To appear}.
\newblock


\bibitem[He et~al.(2022)]%
        {tqp}
\bibfield{author}{\bibinfo{person}{Dong He}, \bibinfo{person}{Supun~C
  Nakandala}, \bibinfo{person}{Dalitso Banda}, \bibinfo{person}{Rathijit Sen},
  \bibinfo{person}{Karla Saur}, \bibinfo{person}{Kwanghyun Park},
  \bibinfo{person}{Carlo Curino}, \bibinfo{person}{Jes\'{u}s
  Camacho-Rodr\'{\i}guez}, \bibinfo{person}{Konstantinos Karanasos}, {and}
  \bibinfo{person}{Matteo Interlandi}.} \bibinfo{year}{2022}\natexlab{}.
\newblock \showarticletitle{{Query Processing on Tensor Computation Runtimes}}.
\newblock \bibinfo{journal}{\emph{Proc. VLDB Endow.}} \bibinfo{volume}{15},
  \bibinfo{number}{11}, \bibinfo{pages}{2811–2825}.
\newblock
\showISSN{2150-8097}


\bibitem[{Heavy.ai}(2022)]%
        {heavy}
\bibfield{author}{\bibinfo{person}{{Heavy.ai}}.}
  \bibinfo{year}{2022}\natexlab{}.
\newblock \bibinfo{title}{{HeavyDB}}.
\newblock \bibinfo{howpublished}{\url{https://github.com/heavyai/heavydb}}.
\newblock


\bibitem[Hu et~al.(2022)]%
        {TCU}
\bibfield{author}{\bibinfo{person}{Yu-Ching Hu}, \bibinfo{person}{Yuliang Li},
  {and} \bibinfo{person}{Hung-Wei Tseng}.} \bibinfo{year}{2022}\natexlab{}.
\newblock \showarticletitle{TCUDB: Accelerating Database with Tensor
  Processors}. In \bibinfo{booktitle}{\emph{SIGMOD 2022}}.
  \bibinfo{pages}{1360–1374}.
\newblock


\bibitem[Huang and Chen(2022)]%
        {joinproject3}
\bibfield{author}{\bibinfo{person}{Zichun Huang} {and} \bibinfo{person}{Shimin
  Chen}.} \bibinfo{year}{2022}\natexlab{}.
\newblock \showarticletitle{Density-Optimized Intersection-Free Mapping and
  Matrix Multiplication for Join-Project Operations}.
\newblock \bibinfo{journal}{\emph{VLDB Endowment}} \bibinfo{volume}{15},
  \bibinfo{number}{10}, \bibinfo{pages}{2244–2256}.
\newblock


\bibitem[Hutchison et~al.(2017)]%
        {lara}
\bibfield{author}{\bibinfo{person}{Dylan Hutchison}, \bibinfo{person}{Bill
  Howe}, {and} \bibinfo{person}{Dan Suciu}.} \bibinfo{year}{2017}\natexlab{}.
\newblock \showarticletitle{{LaraDB: A Minimalist Kernel for Linear and
  Relational Algebra Computation}}. In \bibinfo{booktitle}{\emph{Proceedings of
  the 4th ACM SIGMOD Workshop on Algorithms and Systems for MapReduce and
  Beyond}} \emph{(\bibinfo{series}{BeyondMR'17})}.
\newblock


\bibitem[Kimball and Ross(2013)]%
        {kimball2011data}
\bibfield{author}{\bibinfo{person}{Ralph Kimball} {and} \bibinfo{person}{Margy
  Ross}.} \bibinfo{year}{2013}\natexlab{}.
\newblock \bibinfo{booktitle}{\emph{{The Data Warehouse Toolkit: The Definitive
  Guide to Dimensional Modeling}} (\bibinfo{edition}{3rd} ed.)}.
\newblock \bibinfo{publisher}{Wiley Publishing}.
\newblock
\showISBNx{1118530802}


\bibitem[Koutsoukos et~al.(2021)]%
        {hummingbird2}
\bibfield{author}{\bibinfo{person}{Dimitrios Koutsoukos},
  \bibinfo{person}{Supun Nakandala}, \bibinfo{person}{Konstantinos Karanasos},
  \bibinfo{person}{Karla Saur}, \bibinfo{person}{Gustavo Alonso}, {and}
  \bibinfo{person}{Matteo Interlandi}.} \bibinfo{year}{2021}\natexlab{}.
\newblock \showarticletitle{{Tensors: An Abstraction for General Data
  Processing}}.
\newblock \bibinfo{journal}{\emph{Proc. VLDB Endow.}} \bibinfo{volume}{14},
  \bibinfo{number}{10} (\bibinfo{year}{2021}), \bibinfo{pages}{1797–1804}.
\newblock


\bibitem[Machado et~al.(2022)]%
        {datamesh}
\bibfield{author}{\bibinfo{person}{In\^{e}s~Ara\'{u}jo Machado},
  \bibinfo{person}{Carlos Costa}, {and} \bibinfo{person}{Maribel~Yasmina
  Santos}.} \bibinfo{year}{2022}\natexlab{}.
\newblock \showarticletitle{{Data Mesh: Concepts and Principles of a Paradigm
  Shift in Data Architectures}}.
\newblock \bibinfo{journal}{\emph{Procedia Comput. Sci.}}
  \bibinfo{volume}{196} (\bibinfo{year}{2022}), \bibinfo{pages}{263–271}.
\newblock


\bibitem[Nakandala et~al.(2020)]%
        {hummingbird1}
\bibfield{author}{\bibinfo{person}{Supun Nakandala}, \bibinfo{person}{Karla
  Saur}, \bibinfo{person}{Gyeong-In Yu}, \bibinfo{person}{Konstantinos
  Karanasos}, \bibinfo{person}{Carlo Curino}, \bibinfo{person}{Markus Weimer},
  {and} \bibinfo{person}{Matteo Interlandi}.} \bibinfo{year}{2020}\natexlab{}.
\newblock \showarticletitle{{A Tensor Compiler for Unified Machine Learning
  Prediction Serving}}. In \bibinfo{booktitle}{\emph{OSDI 2020}}.
  \bibinfo{publisher}{USENIX Association}, Article \bibinfo{articleno}{51}.
\newblock


\bibitem[Okuta et~al.(2017)]%
        {cupy}
\bibfield{author}{\bibinfo{person}{Ryosuke Okuta}, \bibinfo{person}{Yuya Unno},
  \bibinfo{person}{Daisuke Nishino}, \bibinfo{person}{Shohei Hido}, {and}
  \bibinfo{person}{Crissman Loomis}.} \bibinfo{year}{2017}\natexlab{}.
\newblock \showarticletitle{{CuPy: A NumPy-Compatible Library for NVIDIA GPU
  Calculations}}. In \bibinfo{booktitle}{\emph{NIPS 2017 Workshop:
  LearningSys}}.
\newblock


\bibitem[O'Neil et~al.(2009)]%
        {ssb}
\bibfield{author}{\bibinfo{person}{Patrick O'Neil}, \bibinfo{person}{Elizabeth
  O'Neil}, \bibinfo{person}{Xuedong Chen}, {and} \bibinfo{person}{Stephen
  Revilak}.} \bibinfo{year}{2009}\natexlab{}.
\newblock \showarticletitle{{The Star Schema Benchmark and Augmented Fact Table
  Indexing}}. In \bibinfo{booktitle}{\emph{Performance Evaluation and
  Benchmarking: First TPC Technology Conference, TPCTC 2009}}.
  \bibinfo{publisher}{Springer-Verlag}, \bibinfo{pages}{237–252}.
\newblock


\bibitem[Park et~al.(2022)]%
        {raven}
\bibfield{author}{\bibinfo{person}{Kwanghyun Park}, \bibinfo{person}{Karla
  Saur}, \bibinfo{person}{Dalitso Banda}, \bibinfo{person}{Rathijit Sen},
  \bibinfo{person}{Matteo Interlandi}, {and} \bibinfo{person}{Konstantinos
  Karanasos}.} \bibinfo{year}{2022}\natexlab{}.
\newblock \showarticletitle{{End-to-End Optimization of Machine Learning
  Prediction Queries}}. In \bibinfo{booktitle}{\emph{SIGMOD 2022}}.
  \bibinfo{pages}{587–601}.
\newblock


\bibitem[Paszke et~al.(2019)]%
        {pytorch}
\bibfield{author}{\bibinfo{person}{Adam Paszke}, \bibinfo{person}{Sam Gross},
  \bibinfo{person}{Francisco Massa}, \bibinfo{person}{Adam Lerer},
  \bibinfo{person}{James Bradbury}, \bibinfo{person}{Gregory Chanan},
  \bibinfo{person}{Trevor Killeen}, \bibinfo{person}{Zeming Lin},
  \bibinfo{person}{Natalia Gimelshein}, \bibinfo{person}{Luca Antiga},
  {et~al\mbox{.}}} \bibinfo{year}{2019}\natexlab{}.
\newblock \showarticletitle{{Pytorch: An imperative style, high-performance
  deep learning library}}. In \bibinfo{booktitle}{\emph{NeurIPS 2019}},
  Vol.~\bibinfo{volume}{32}.
\newblock


\bibitem[Psallidas et~al.(2022)]%
        {tree_pop}
\bibfield{author}{\bibinfo{person}{Fotis Psallidas}, \bibinfo{person}{Yiwen
  Zhu}, \bibinfo{person}{Bojan Karlas}, \bibinfo{person}{Jordan Henkel},
  {et~al\mbox{.}}} \bibinfo{year}{2022}\natexlab{}.
\newblock \showarticletitle{{Data Science Through the Looking Glass: Analysis
  of Millions of GitHub Notebooks and ML.NET Pipelines}}.
\newblock \bibinfo{journal}{\emph{SIGMOD Rec.}} \bibinfo{volume}{51},
  \bibinfo{number}{2} (\bibinfo{year}{2022}), \bibinfo{pages}{30–37}.
\newblock
\showISSN{0163-5808}
\urldef\tempurl%
\url{https://doi.org/10.1145/3552490.3552496}
\showDOI{\tempurl}


\bibitem[{Rapidsai}(2022)]%
        {cudf}
\bibfield{author}{\bibinfo{person}{{Rapidsai}}.}
  \bibinfo{year}{2022}\natexlab{}.
\newblock \bibinfo{title}{{cuDF}}.
\newblock \bibinfo{howpublished}{\url{https://github.com/rapidsai/cudf}}.
\newblock


\bibitem[Sun et~al.(2023)]%
        {wenbo_hard}
\bibfield{author}{\bibinfo{person}{Wenbo Sun}, \bibinfo{person}{Asterios
  Katsifodimos}, {and} \bibinfo{person}{Rihan Hai}.}
  \bibinfo{year}{2023}\natexlab{}.
\newblock \showarticletitle{{An Empirical Performance Comparison between Matrix
  Multiplication Join and Hash Join on GPUs}}. In
  \bibinfo{booktitle}{\emph{ICDE 2023 Workshop: HardBD {\&} Active}}.
  \bibinfo{pages}{To appear}.
\newblock


\bibitem[{Transaction Processing Performance Council}(2018)]%
        {tpch}
\bibfield{author}{\bibinfo{person}{{Transaction Processing Performance
  Council}}.} \bibinfo{year}{2018}\natexlab{}.
\newblock \bibinfo{title}{{TPC Benchmark H.}}
\newblock
  \bibinfo{howpublished}{\url{http://tpc.org/tpc_documents_current_versions/pdf/tpc-h_v2.18.0.pdf}}.
\newblock


\bibitem[Yuster and Zwick(2005)]%
        {sparse}
\bibfield{author}{\bibinfo{person}{Raphael Yuster} {and} \bibinfo{person}{Uri
  Zwick}.} \bibinfo{year}{2005}\natexlab{}.
\newblock \showarticletitle{{Fast Sparse Matrix Multiplication}}.
\newblock \bibinfo{journal}{\emph{ACM Trans. Algorithms}} \bibinfo{volume}{1},
  \bibinfo{number}{1} (\bibinfo{year}{2005}), \bibinfo{pages}{2–13}.
\newblock
\showISSN{1549-6325}
\urldef\tempurl%
\url{https://doi.org/10.1145/1077464.1077466}
\showDOI{\tempurl}


\end{thebibliography}

\end{document}